\documentclass[structabstract]{aa}
\usepackage{graphicx}
\usepackage{natbib}
\usepackage{lscape}
\usepackage{longtable}
\newcommand{\lsim}{\ \raise -2.truept\hbox{\rlap{\hbox{$\sim$}}\raise
5.truept\hbox{$<$}\ }}
\newcommand{\gsim}{\ \raise -2.truept\hbox{\rlap{\hbox{$\sim$}}\raise
5.truept\hbox{$>$}\ }}
\newcommand{\nodata}{...}
\newcommand{\dplz}{$D_{PLZ}$\ }
\newcommand{\dpl}{$D_{PL}$\ }
\newcommand\mM{\ifmmode(m{-}M)\else$(m{-}M)$\fi}
\begin{document}
   \title{The distance to NGC\,1316 (Fornax~A): yet another curious
     case\footnote{Based on data obtained from the ESO Science Archive
       Facility, and on observations made with the NASA/ESA Hubble
       Space Telescope, and obtained from the Hubble Legacy Archive,
       which is a collaboration between the Space Telescope Science
       Institute (STScI/NASA), the Space Telescope European
       Coordinating Facility (ST-ECF/ESA) and the Canadian Astronomy
       Data Centre (CADC/NRC/CSA).}}

   \subtitle{}

   \author{Michele Cantiello\inst{1}
          \and
          Aniello Grado\inst{2}
      \and
      John P. Blakeslee\inst{3}
      \and
      Gabriella Raimondo\inst{1}
       \and
     Gianluca Di Rico\inst{1}
      \and
      Luca Limatola\inst{2}
      \and
      Enzo Brocato\inst{1,4}
      \and
      Massimo Della Valle\inst{2,5}
      \and
      Roberto Gilmozzi\inst{6}
}
\institute{INAF Osservatorio Astronomico di Teramo, via M. Maggini
  snc, I-64100, Teramo, Italy  \\ \email{cantiello@oa-teramo.inaf.it}
         \and
INAF Osservatorio Astronomico di Capodimonte, salita Moiariello I-80131 Napoli, Italy \\
        \and
Dominion Astrophysical Observatory, Herzberg Institute of Astrophysics,
          National Research Council of Canada, Victoria, Canada \\
 \and
INAF Osservatorio Astronomico di Roma, Via Frascati 33, I-00040, Monte
Porzio Catone, Roma, Italy \\
 \and
 International Centre for Relativistic Astrophysics, Piazzale della Repubblica 2, I-65122, Pescara, Italy \\
\and
 European Southern Observatory, Karl-Schwarzschild-Str. 2, 85748 Garching bei M\"unchen, Germany 
          }
   \date{Received --; accepted --}

\authorrunning{Cantiello et al.}
\titlerunning{The distance of NGC\,1316}


  \abstract
   {}
  {The distance of NGC\,1316, the brightest galaxy in the Fornax cluster, provides an interesting test for the cosmological distance scale. First, because
    Fornax is the second largest cluster of galaxies within
    $\lsim25$ Mpc after Virgo and, in contrast to Virgo, 
    has a small line-of-sight depth; and second, because NGC\,1316 is the single
    galaxy with the largest number of detected Type Ia supernovae (SNe~Ia),
    giving the opportunity to test the consistency of SNe~Ia distances
    both internally and against other distance indicators.}
   {We measure surface brightness fluctuations (SBF) in NGC\,1316 from ground- and space-based imaging data. The sample provides a homogeneous set of measurements over a wide wavelength interval. The SBF magnitudes, coupled with empirical and theoretical absolute SBF calibrations, are used to estimate the distance to the galaxy. We also present the first $B$-band SBF measurements of NGC\,1316 and use them together with the optical and near-IR SBF data to analyze the properties of field stars in the galaxy.}
   {We obtain $\mM=31.59\pm0.05({\rm stat.})\pm0.14({\rm sys.)}$ mag, or $d=20.8\pm0.5({\rm stat.})\pm1.5({\rm sys.)}$ Mpc. When placed in a consistent Cepheid distance scale, our result agrees with the distances from other indicators. On the other hand, our distance is $\sim17\%$ larger than the most recent estimate based on SNe~Ia. Possible explanations for this disagreement are the uncertain level of internal extinction, and/or calibration issues. Concerning the stellar population analysis, we confirm the results from other spectro-photometric indicators: the field stars in NGC\,1316 are dominated by a component with roughly solar metallicity and intermediate age. A non-negligible mismatch exists between $B$-band SBF models and data. We confirm that such behavior can be accounted for by an enhanced percentage of hot horizontal branch stars.}
  {Our study of the SBF distance to NGC\,1316, and the comparison with distances from other indicators, raises some concern about the homogeneity between the calibrations of different indicators. If not properly placed in the same reference scale, significant differences can occur, with dramatic impact on the cosmological distance ladder.  Our results on the stellar populations properties show that SBF data over a broad wavelength interval are an efficient means of studying the properties of unresolved systems in peculiar cases like NGC\,1316.}

   \maketitle


\section{Introduction}

The past three decades have seen remarkable progress in the study of
the distance scale of the Universe
\citep[][]{jacoby92,ferrarese00,freedman01,freedman10}, resulting in a
general convergence of distances based on different
indicators. Nevertheless, to resolve the lingering
discrepancies \citep{tammann08,freedman10}, there is an urgent need to
$(i)$ lower the statistical (intrinsic) and systematic (external)
errors for single distance indicators; $(ii)$ obtain distance
measurements from indicators with a large range of applicability in
terms of distances and of useful targets in order to minimize the
error propagation over the cosmological distance scale; and $(iii)$
improve/analyze the matching between independent indicators. A particularly
promising distance indicator for addressing the three listed items
is the surface brightness fluctuations technique
\citep[SBF,][]{ts88,tal90,blake09}.

After the first two ``rungs'' of the cosmological distance scale,
represented by geometric methods and by primary indicators (variable
stars, main-sequence fitting, etc.), the SBF method is one of the
most accurate indicators, with a median 0.14 mag accuracy on distance
moduli, or $\sim 7$ \% in distance, up to $\sim100$ Mpc
\citep{mei03,biscardi08}.  However, the accuracy gets considerably
smaller, to a mean $\sim0.08$ mag ($\sim$4\% in distance) from
space-based optical data (Table \ref{tab_accuracy})

By definition, the SBF signal corresponds to the ratio of the second
to the first moments of the luminosity function of stars in a
galaxy.  As opposed to the surface brightness that does not
scale with the distance of the galaxy, the SBF signal scales inversely
with the distance squared. 
Observationally, the SBF method relies on the measurement of the
intrinsic flux variance in a galaxy, generated by the Poissonian
fluctuations in the surface brightness due to the statistical variation
of the stellar counts in adjacent resolution elements. The variance,
normalized to the local mean surface brightness, is converted to an
apparent magnitude, $\bar{m}$, from which the distance modulus,
$\bar{m}-\bar{M}$, follows once the absolute $\bar{M}$ is known.

Given its definition, $\bar{M}$ in a given bandpass is 
dependent on the properties of the underlying stellar populations.
The analysis of large samples of early-type systems, including galaxy
in groups, has made it possible to characterize the dependence of $\bar{M}$ 
on stellar population properties using linear relations with respect to
some broad-band optical colour \citep{tonry01,mei07xiii,cantiello07b}.

In Table \ref{tab_accuracy} we report the median errors on SBF
measurements, $\delta(\bar{m})$, and on the associated distance
moduli, $\delta(\bar{m}{-}\bar{M})$, derived from different samples.
For optical SBF, the typical accuracy for ground-based measurements
has been $\lsim$0.2 mag (10\% in distance). Because this is dominated
by measurement errors, the superior resolution of HST leads to an
improved accuracy of $\sim$0.08 mag. For a comparison, the internal
scatter of the period-luminosity (PL) relation of Cepheids ranges from
$0.20$ mag in the $V$ band, to $0.09$ mag in the Spitzer [3.6] $\mu m$
and [4.5] $\mu m$ bands \citep[e.g.,][]{ngeow09}. The mean
uncertainties for the near-IR SBF sample quoted in Table
\ref{tab_accuracy} are generally larger than the optical ones,
$\lsim$0.2 mag ($\sim$10\% on distances) even in the case of
space-based data. This is due both to observational/technical issues
\citep[e.g. dark current patterns, ``wormy'' background,
  see][]{jensen98,jensen01}, and to the scatter of the calibration
\citep[mostly related to the sensitivity of the SBF signal to the
  properties of AGB and TP-AGB stars in these bands,
  see][]{mei01,liu02,jensen03,raimondo05,gonzalez10}. The availability
of Wide Field Camera~3 (WFC3, on board of the {\it Hubble Space
  Telescope}, HST) is expected to significantly improve the situation
in this wavelength regime.  Finally, the zero point of the calibration
is typically tied to the Cepheid distance scale to an accuracy $\lsim$
0.08 mag \citep{tonry00,blake10b}.

Thus, SBF distances are characterized by reasonably well defined
and small (especially in optical bands) internal and calibration
errors (item $(i)$ in the list above). The technique has been used to
estimate distances for Local Group galaxies \citep[even closer than
  that if one takes into account the work on Galactic globular
  clusters by][]{ajhar94}, out to galaxies at $\gsim100$ Mpc
\citep{jensen01,biscardi08}. With the highly improved near-IR imaging
capabilities of the WFC3/IR and similar instruments, and
thanks to the much brighter SBF signal in the near-IR, the upper limit
on SBF distances is expected to increase significantly. Thus, SBF can
encompass more than two orders of magnitude in distance, bridging
local to cosmological distances with the use of a single indicator
(item $(ii)$ above).

In this paper we present measurements of SBF magnitudes for the
intermediate-age merger remnant \object{NGC\,1316}, also known as Fornax~A
\citep[e.g.,][]{schweizer80,terlevich02}. This galaxy is peculiar in
many ways.  Although by far the brightest member of the Fornax
cluster, it is not near the cluster center, being $\sim3.7^{\circ}$
away from the central giant elliptical \object{NGC\,1399} (projected separation
of $\sim1.3$ Mpc at the distance of Fornax). It shows numerous dust
features, a prominent dust lane, $H\alpha$ filaments, loops, and tidal
tails originally analyzed by \citet{schweizer80}. Moreover, it is a
powerful radio-galaxy and, to date NGC\,1316 is the single galaxy with
the largest number of discovered Type Ia supernovae (SNe~Ia hereafter;
four events recorded). The latter property makes NGC\,1316 a
remarkable place to test the extragalactic distance scale, because of
the role of SNe~Ia for cosmological distances, and because NGC\,1316
is one of the nearest massive post-merger galaxies.

We have collected data covering a large
wavelength interval (from $B$ to $H$ band), with the specific purpose
of performing a self-consistent analysis of SBF data for this galaxy in
order to carry out a comprehensive study of the SBF in NGC\,1316, and
consequently, of the galaxy distance (item $(iii)$ above).

Furthermore, given their dependence on the square of the stellar
luminosity, SBF magnitudes are especially sensitive to the brightest
stars at a particular wavelength and at a given evolutionary phase of
a stellar population.  As shown by various authors SBF magnitudes and,
in particular, SBF colours, can be used to investigate the properties
of a specific stellar component in the host stellar population,
depending on the observing wavelength
\citep{worthey93b,bva01,cantiello03,jensen03,raimondo05}.  As an
example, SBF colours involving bluer bands, like $B$, have been used
to study the hot stellar component in unresolved systems
\citep[e.g.,][]{cantiello07a}. Likewise, specific phenomena like the
mass-loss rates in the AGB phase have been analyzed by taking advantage
of near-IR SBF data \citep{raimondo09,gonzalez10}.  As a consequence
of the quoted relation between the SBF signal and stellar population
properties, we take advantage of the broad passband coverage to
characterize the properties of field stars in the galaxy, in order
also to provide new constraints on the formation history and
evolution of the peculiar galaxy NGC\,1316.

\begin{table*}
\caption{Median $\bar{m}$ and $(\bar{m}{-}\bar{M})$ errors from the
literature}
\label{tab_accuracy}
\centering
\begin{tabular}{l c c l c }
\hline\hline 
Sample               & $\delta(\bar{m})$ & $\delta(\bar{m}{-}\bar{M})$  & Filter         & Number of sources  \\
                     &   (mag)           &       (mag)                  &                & (space/ground obs.) \\
\hline
\multicolumn{5}{c}{Optical bands} \\
\citet{tonry01}      &   0.18  &    0.20  &      $I$                  &      280 (ground)    \\
\citet{cantiello07b} &   0.02  &    0.09  & $F814W$($\sim I$)        &       13 (space) \\
\citet{blake09}      &   0.04  &    0.08  & $F850LP$($\sim g_{SDSS}$)  &      134 (space) \\
\citet{blake10b}     &   0.02  &    0.07  &  $F814W$($\sim I$)       &        9 (space) \\
\hline
\multicolumn{5}{c}{Near-IR bands} \\
\citet{jensen98}     &   0.14  &    0.19  & $K'$                    &       16   (ground)   \\
\citet{liu02}        &   0.07  &    0.20  & $K_S$                    &       19  (ground)   \\
\citet{jensen03}     &   0.08  &    0.17  & $F160W$ ($\sim H$)       &       79 (space) \\
\hline\hline 
\end{tabular} \\
\end{table*}

\noindent The organization of the paper is as follows. In \S \ref{sec_data} we
present a description of the imaging data used, the data reduction and
calibration procedures. Sections \S \ref{sec_measure} and \S
\ref{sec_distances} describe the procedures for SBF measurements and
the calibrations used to determine the distances.  In section \S
\ref{sec_compare} we compare the SBF distance to previous results from
other indicators. We analyze the properties of the unresolved field
star populations in \S \ref{sec_ssp}, and summarize our conclusions
in \S \ref{sec_conclu}. In Appendix \ref{appenda} a detailed
comparison of SBF and PNLF distances is presented. Finally, Appendix
\ref{appendb} presents some details on the SBF versus SNe~Ia
comparison.


\section{Observations and reductions}
\label{sec_data}

\begin{table*}
\caption{Main properties of NGC\,1316.}
\centering
\begin{tabular}{l c}
\hline\hline
Alternative names                      &  \object{Fornax A}, \object{FCC\,21}, \object{Arp\,154}, \object{ESO\,357-G\,022} \\
RA(J2000)$^{\mathrm{1}}$                   &   03h22m41.7s      \\
Dec(J2000)$^{\mathrm{1}}$                  &   $-$37d12m30s       \\
Galaxy Type$^{\mathrm{2}}$	                &   S0		     \\
Morphological Type$^{\mathrm{2}}$          &  $-$1.8$\pm$0.7	     \\
Absolute $B$-band magnitude$^{\mathrm{2}}$ &	$-$22.5 mag       \\
Recorded SNe Ia events	                       & \object{SN\,1980N},\object{SN\,1981D}, \object{SN\,2006dd}, \object{SN\,2006mr} \\
$cz^{\mathrm{1}}$ (km/s, Heliocentric)    &  1788$\pm$10  	     \\
$E(B{-}V)^{\mathrm{3}}$                   &  0.021 mag    	     \\
$(V{-}I)_0^{\mathrm{4}}$		       & 1.132 $\pm$ 0.016 mag  \\
\hline
\end{tabular}
\begin{list}{}{}
\item[$^{\mathrm{1}}$] Data retrieved from NED (http://nedwww.ipac.caltech.edu); 
$^{\mathrm{2}}$  Hyperleda (http://leda.univ-lyon1.fr);
$^{\mathrm{3}}$ \citet{sfd98};
$^{\mathrm{4}}$  \citet{tonry01}
\end{list}
\label{tab_obs}
\end{table*}

This work is based on data of NGC\,1316 from the {\it Very Large
  Telescope} (VLT) and HST archives. We used $i)$ $B$, $V$ and
$I$-band observations obtained with the FORS1 Imager at ESO's
VLT in Paranal (Program 64.H-0624(A), P.I. M. Della Valle), and from
the HST $ii)$ ACS $F475W$ and $F850LP$-band data from the ACSFCS
survey \citep{jordan07}, $iii)$ WFC3/IR observations in the $F110W$
and $F160W$ filters, plus WFC3/UVIS $F336W$ (HST Program ID 11691,
P.I. P. Goudfrooij).

It is useful to note, for the forthcoming discussion, that the VLT
observations were part of a project aimed at discovering, monitoring
and characterizing the properties of the nova population in NGC\,1316,
with the specific purpose of deriving the distance to the galaxy using
novae \citep{dellavalle02}. Some relevant properties of the target
are listed in Table \ref{tab_obs}.

In the remainder of this section we describe the reduction and
calibration procedures adopted for the data from both telescopes. In
all cases we used SExtractor \citep{bertin96} for the source
photometry, and the IRAF/STSDAS task ELLIPSE \citep[based on the
  method described by][]{jedrzejewski87} to fit the galaxy isophotes.

\subsection{VLT data}
\label{sec_reduc}

   \begin{figure*}
   \centering
   \includegraphics[width=0.48\textwidth]{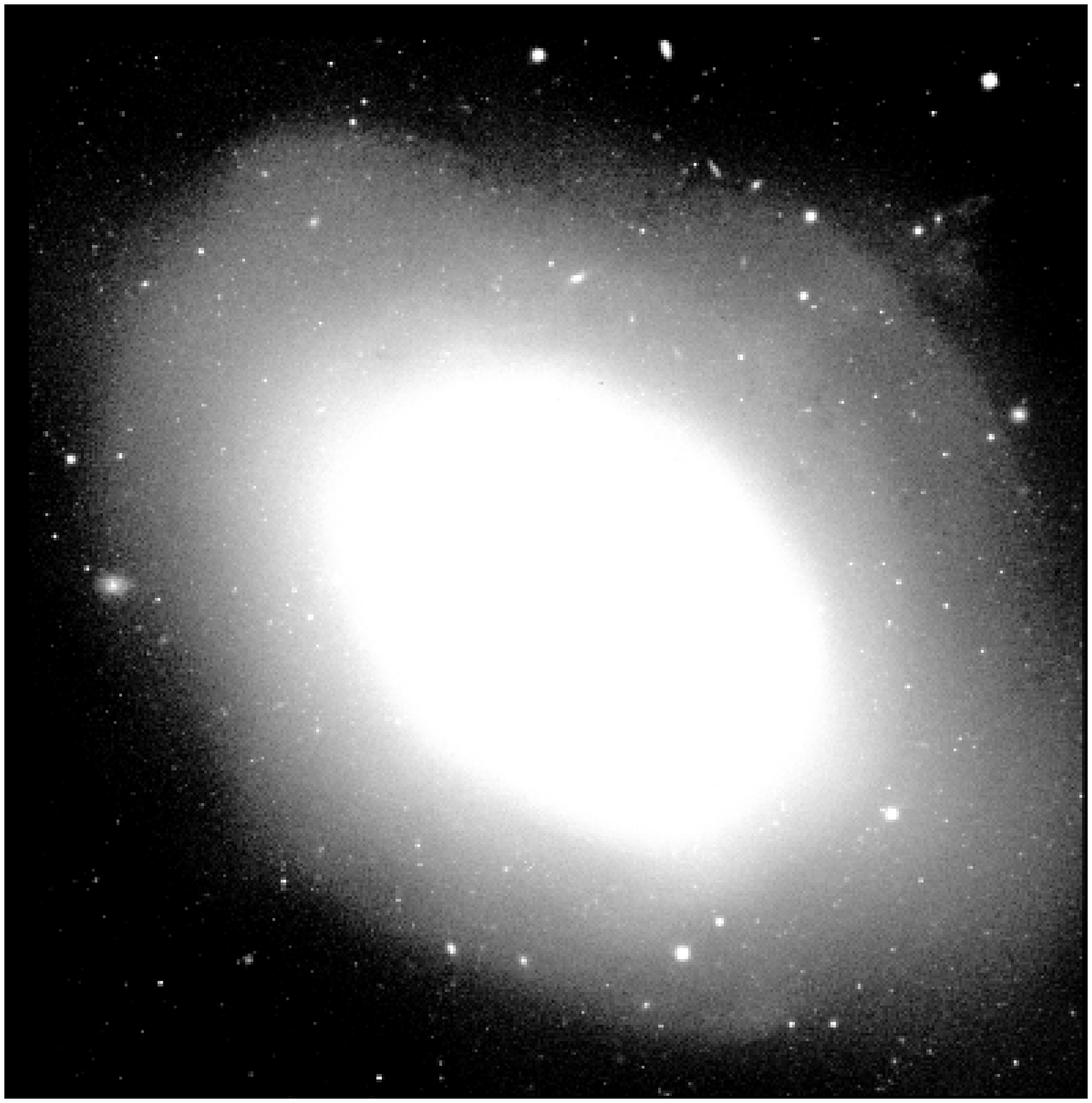}
   \includegraphics[width=0.48\textwidth]{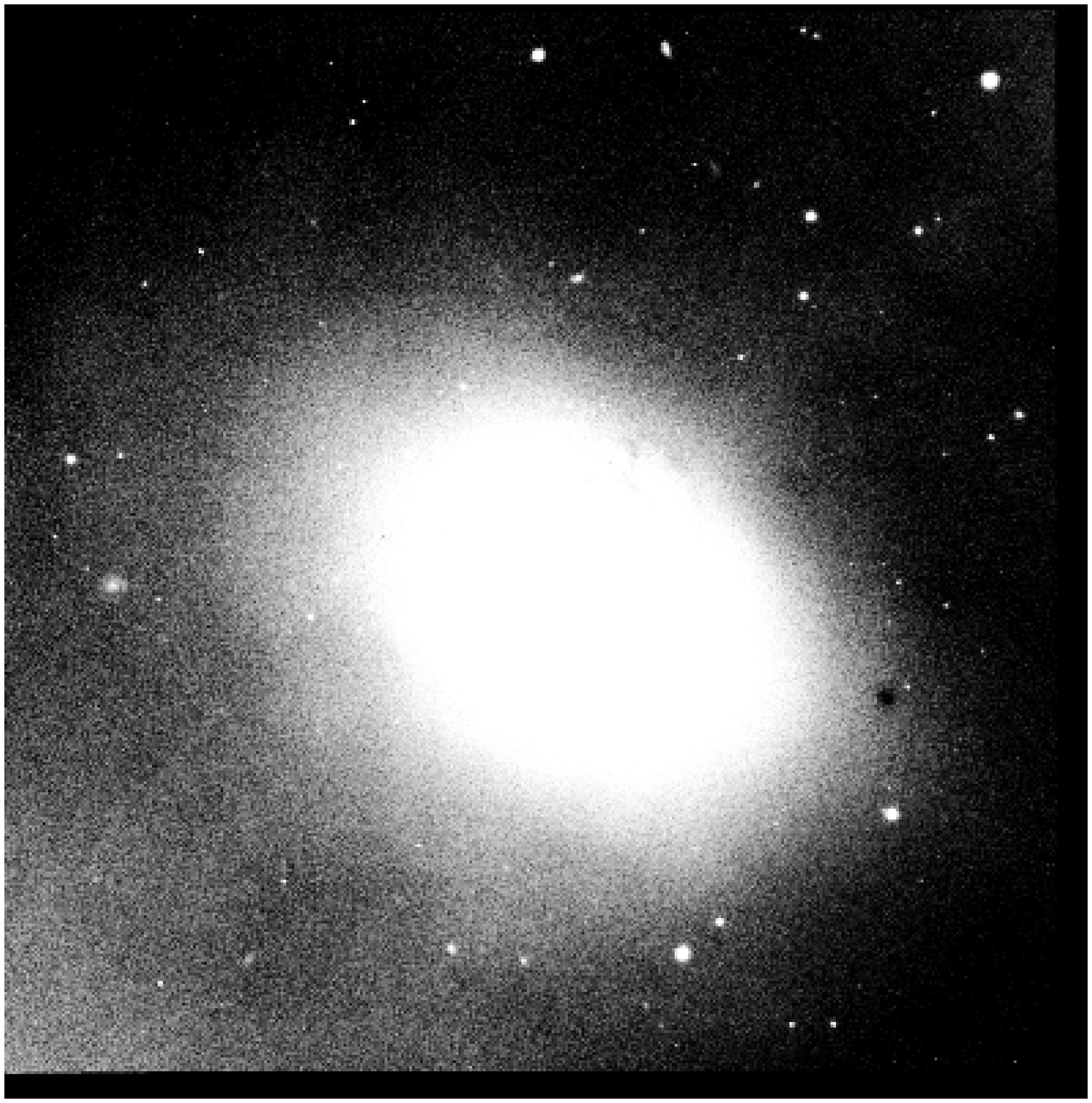}
   \caption{Left panel: Mosaic single exposure in $B$ on 1999 December
     27$^{th}$ (600s total, the first observing run of the
     proposal). Right panel: As left, but for an exposure taken during
     the night of 2000 January 20$^{th}$.}
   \label{fig1}
   \end{figure*}
%

   \begin{table*}
   \begin{center}
      \caption[]{Total and used exposure times for each band.}
         \label{ObsLog}
         \begin{tabular}{lccc}
            \hline\hline
            \noalign{\smallskip}
          &  $B$ (s) & $V$ (s) &  $I$ (s)\\
            \hline 
            \noalign{\smallskip}
            \noalign{\smallskip}
        Total available &  13200 & 9000  & 10800  \\
        Used  &    6000 & 3600  & 6000   \\
             \noalign{\smallskip}
            \hline
	\end{tabular}
	\end{center}
   
   \end{table*}

\subsubsection{Data reduction}

The data reduction was carried out with the VST-Tube imaging pipeline
\citep{grado04,grado12}, specifically developed by one of the
co-authors of this work (A.G.) for data from the VLT Survey Telescope
\citep[VST]{capaccioli05,capaccioli11}. VST-Tube is a very versatile
software for astronomical data analysis, tested against imaging data
taken with different telescopes/detectors, adaptable to existing or
future multi-CCD cameras (more details will be given in a dedicated
forthcoming publication, A.~Grado et~al., in prep.).  Further, 
VST-Tube offers the great advantage of
fully controlling each step of the data processing.

In Table \ref{ObsLog} we report the total exposure times available for
each VLT filter. Unfortunately, technical problems with the camera
made a fraction of the total observing time unusable (55\%, 60\%
  and 44\% of the total exposure time in $B$, $V$, and $I$ band
  respectively). The VLT images downloaded from the archive and used
for this work showed a strong degradation during the thirteen
observing runs spanning nearly two months, from December 1999 to
February 2000 (Figure \ref{fig1}). This is partly due to moon
illumination, and also to the problem of instrument contamination. A
decontamination process of FORS1 was performed in November 1999, and
one more in January 2000 \citep{cavadore99}. The effect of the last
intervention is clearly visible in a $\gsim$1.0 mag jump in the
background level between observations before and after the
decontamination (see Table \ref{QC}). For this reason,
we decided to select only
images taken before 2000 January 15$^{th}$ and, in case of the $I$
filter, we also included the last two exposures of 2000 February
4$^{th}$ (column ``Data Used'' in Table \ref{QC}).  The total exposure
times used after the selection for good frames are also reported in
Table \ref{ObsLog}.

\begin{figure*}
   \centering
    \includegraphics[width=0.3\textwidth]{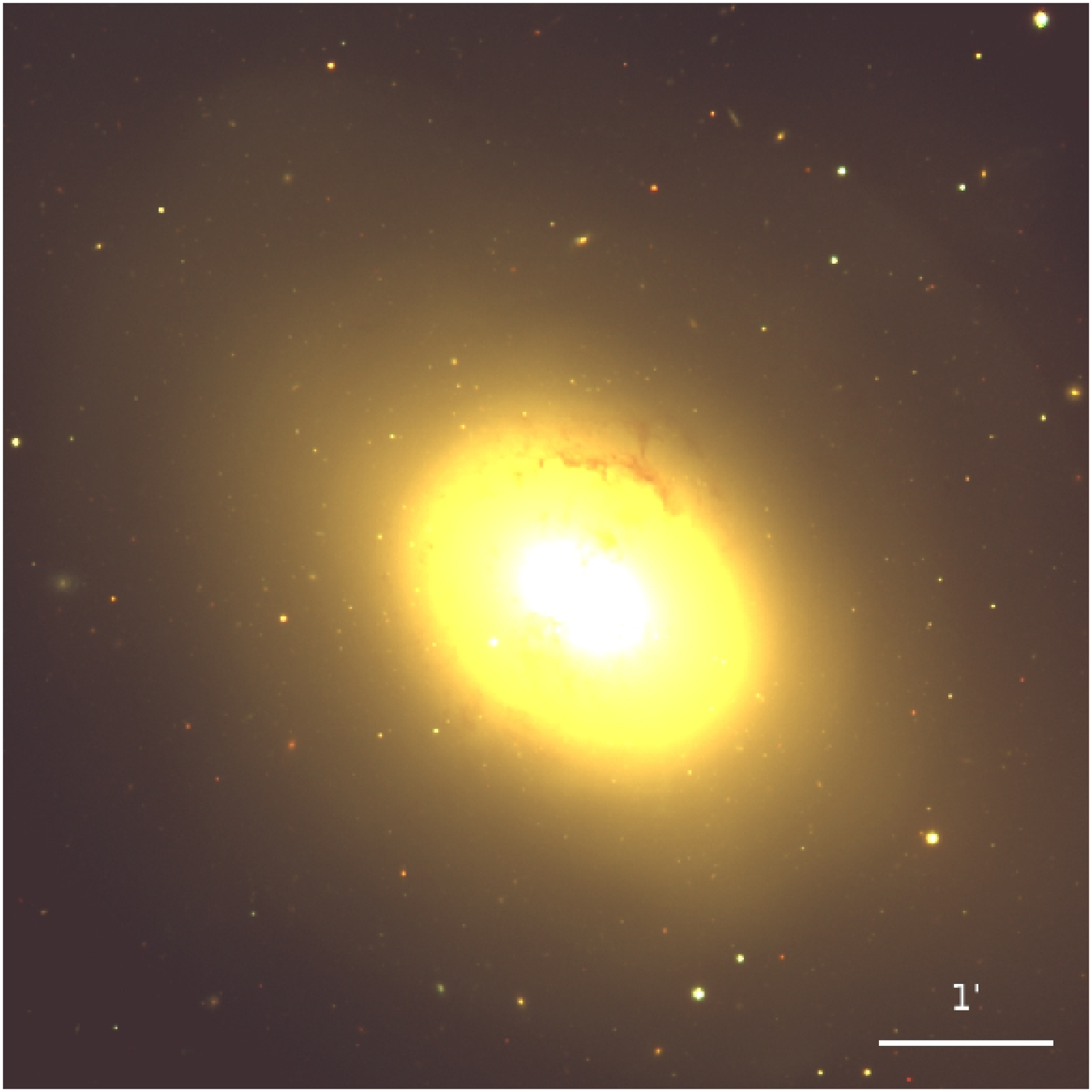} 
  \includegraphics[width=0.3\textwidth]{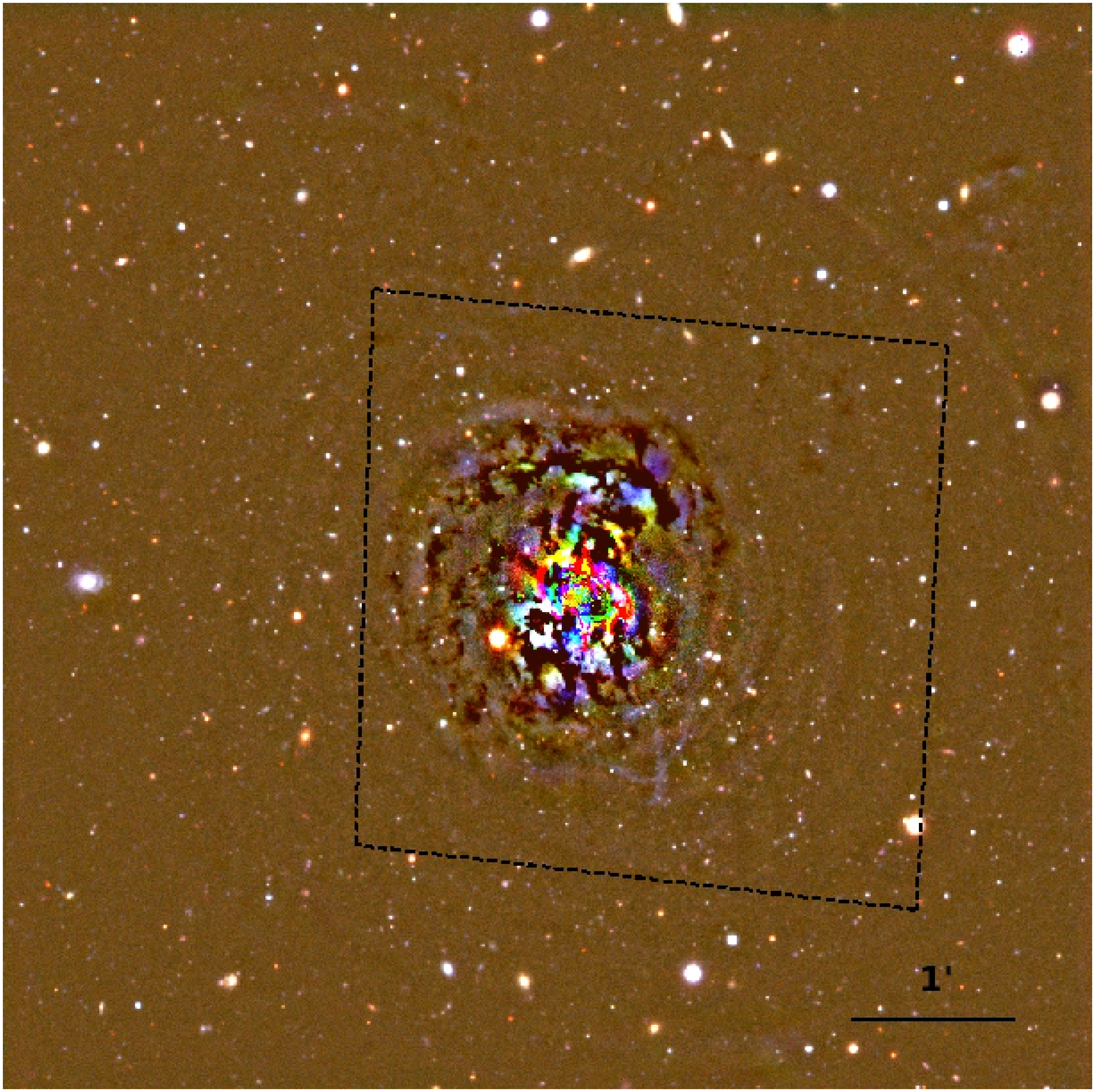} 
   \includegraphics[width=0.3\textwidth]{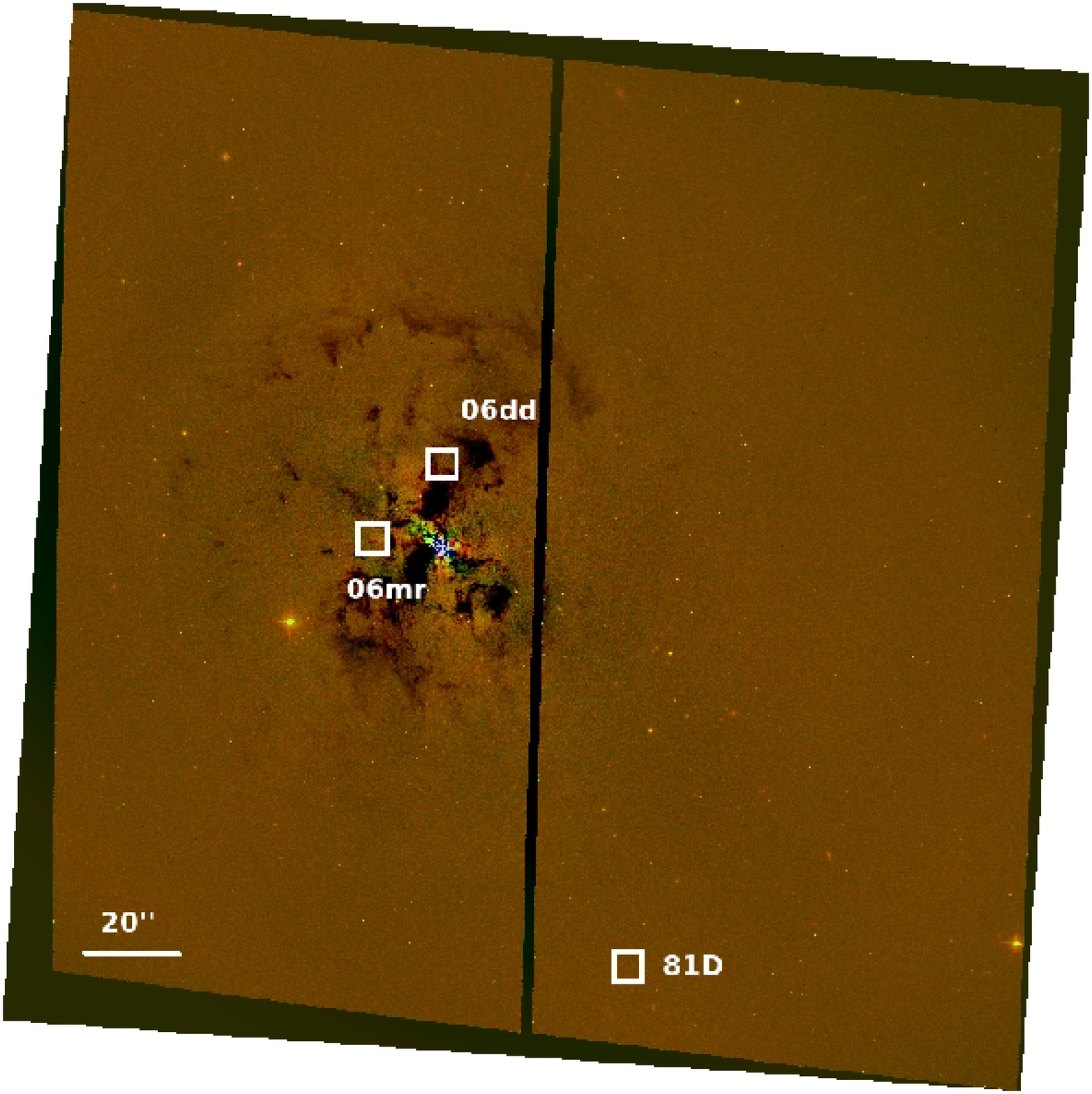} 
     \caption{Left panel: false colour combined VLT $BVI$
         frames. Middle panel: as upper panel, but for residual
         frames. The location of the ACS frames is outlined in black.
       Right panel: ACS and WFC3/UVIS combined residual images. Colour
       coding is chosen to enhance the presence of dust. The sites of
       three over four of the SNe~Ia host by the galaxy are shown.}
         \label{img_res}
   \end{figure*}

The images were reduced as usual removing the instrumental signatures
(overscan correction, bias subtraction, flat field correction and, in
the case of the $I$~band, fringe pattern removal). The resulting co-added
mosaic are approximately $6.9 \times 7.0~arcmin^2$, a colour combined
image of the three mosaics is shown in Figure \ref{img_res} (upper
panel).

\subsubsection{Calibration}

   \begin{figure*}
   \centering
   \includegraphics[width=0.48\textwidth]{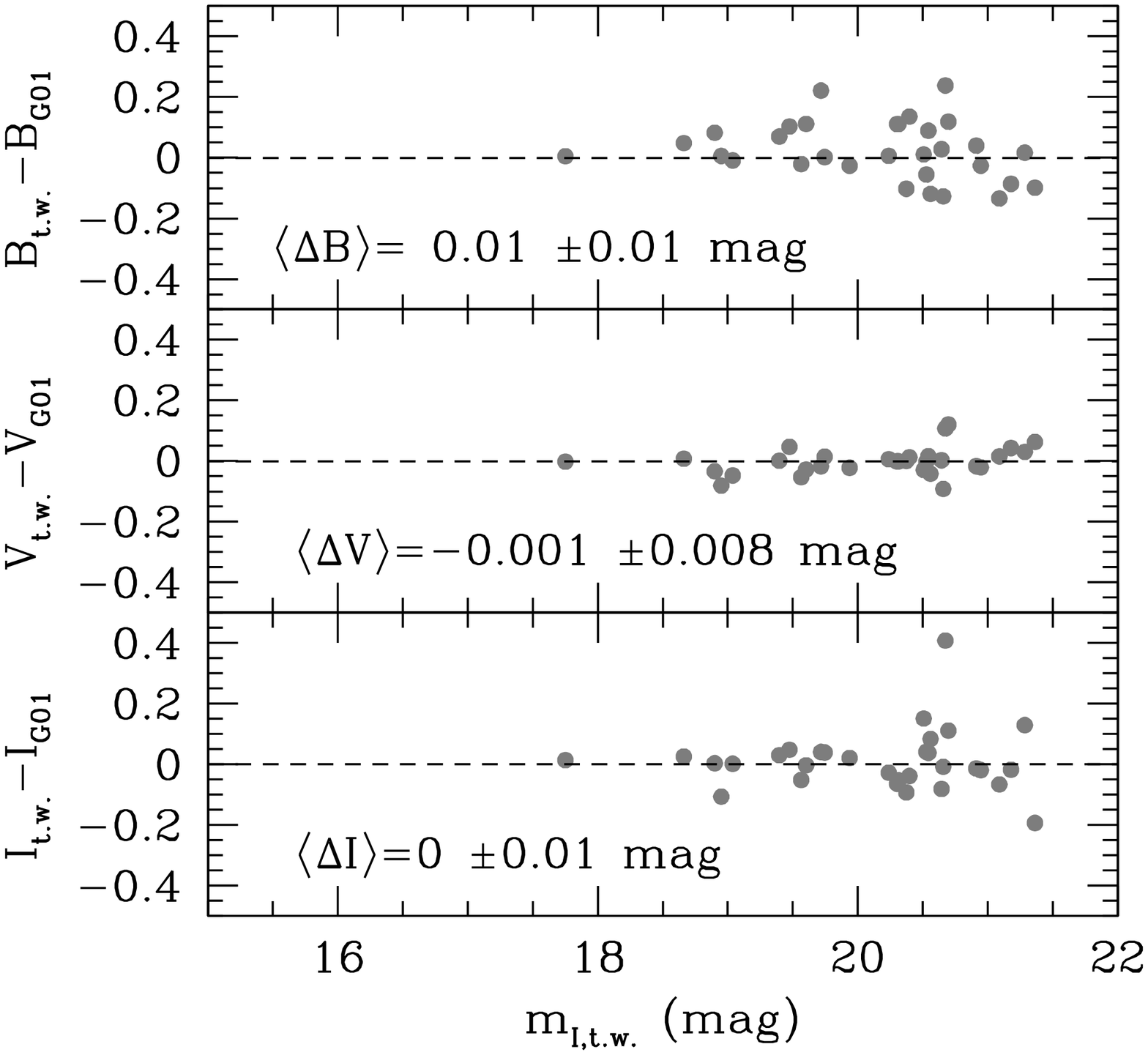} 
   \includegraphics[width=0.48\textwidth]{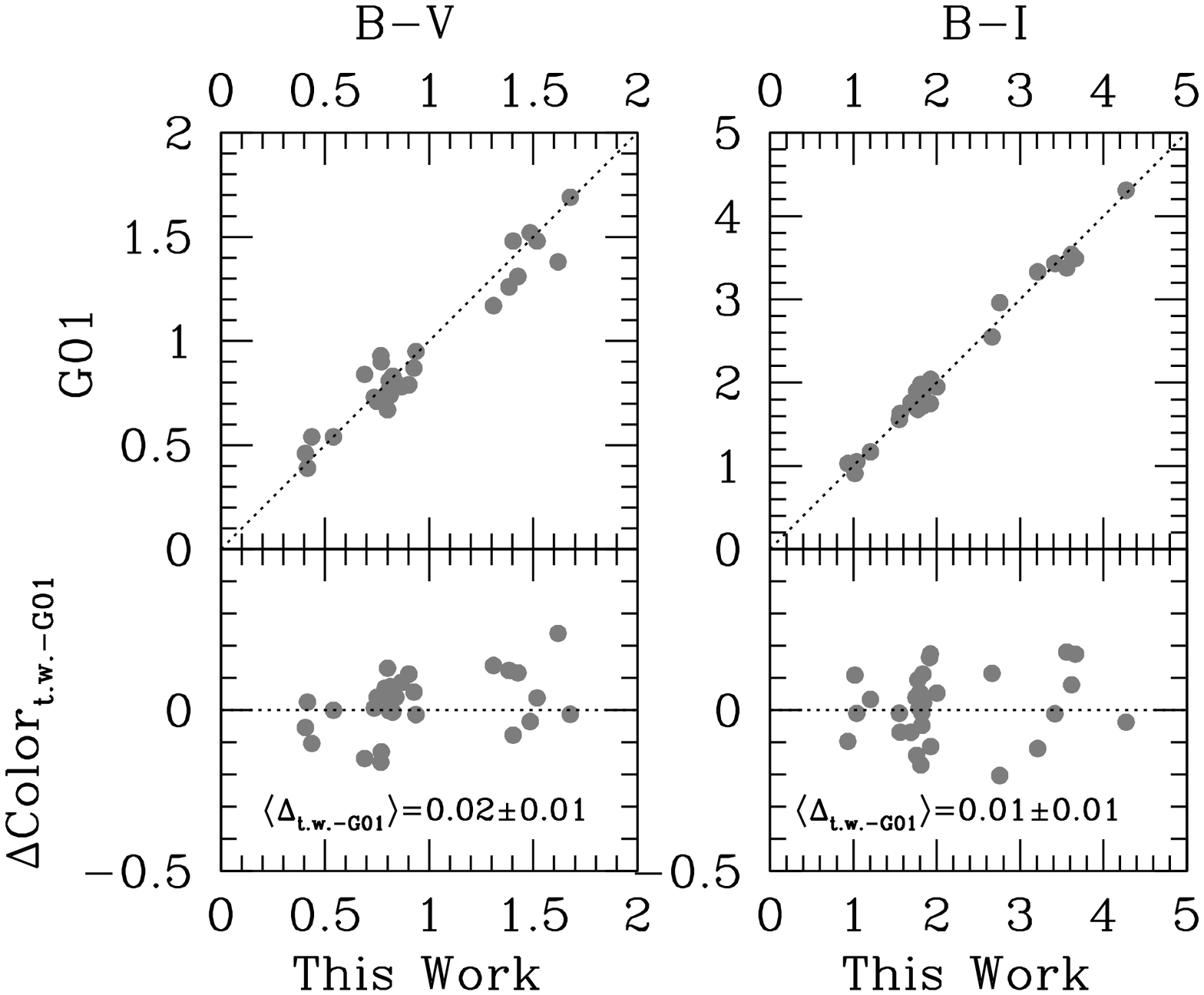} 
      \caption{Comparison between the VLT (this work, t.w.) $BVI$
        photometry and NTT data from G01 used for calibrating FORS1
        photometry.}
         \label{g01}
   \end{figure*}

The VLT data were calibrated adopting the \citet[][G01
  hereafter]{goudfrooij01b} photometry as reference, then solving the
photometric equations accordingly.  G01 obtained optical photometry of
the sources in the field of NGC\,1316 using archival NTT/EMMI data in
$B$, $V$, and $I$ filters. Furthermore, the authors also derived
$JHK_S$ photometry for eight candidate globular clusters (GCs) in the
galaxy, using the IRAC2 camera mounted on the ESO/MPI 2.2m telescope.

The panels in Figure \ref{g01} show the $B$, $V$ and $I$ VLT
photometry versus the G01 calibrating data. Note that G01 adopted
Galactic extinction $E(B-V)=0.0$ from \citet{burstein84}, while we
adopted $E(B-V)=0.021$ from \citet{sfd98}\footnote{The new
  measurements of dust reddening from \citet{sf11} provide $\sim0.002$
  mag smaller Galactic extinction, being $E(B-V)=0.019$. The effect of
  this change is at most $\sim0.02$, for $B$ magnitudes, and
  $\sim0.002$ mag on the colour indices used throughout the present
  work.}.

The comparisons in Figure \ref{g01} are obtained in the same
observational conditions, prior to corrections for Galactic extinction.

\subsection{HST data}

HST instruments, thanks to the high resolution and the sharp PSF,
provide ideal observational datasets for SBF analysis
\citep[e.g.,][]{ajhar97,jensen01,biscardi08,blake10b}.  We analyze
the optical ACS data and provide the first near-IR WFC3/IR SBF
analysis, with the specific purpose of deriving a consistent set of
ground and space-based SBF measurements to secure a reliable distance
to NGC\,1316.

   \begin{figure}
   \centering
   \includegraphics[width=0.48\textwidth]{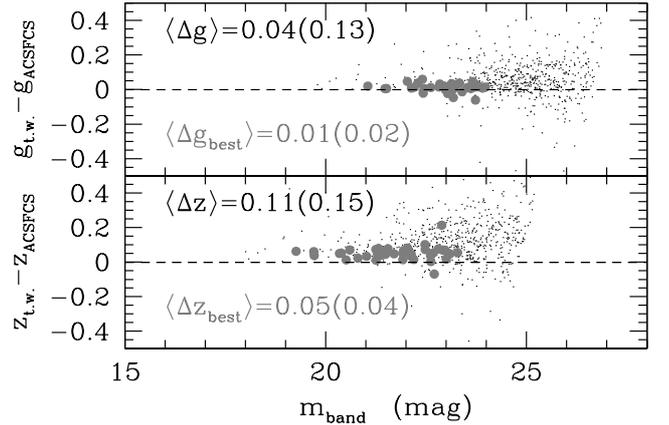}
      \caption{Comparison between our photometry and the ACSFCS
        results. Black dots and labels refer to the whole sample of
        $\sim$620 matched sources, gray circles and labels to the
        $\sim$40 selected sources. The median difference and the
        associated $rms$ are also reported.}
         \label{fcs}
   \end{figure}

\subsubsection{NGC\,1316 as seen by the ACS}
We analyzed the ACS $F475W$ ($\sim$ SDSS $g$ band) and $F850LP$
($\sim$ SDSS $z$) observations of NGC\,1316 obtained for the ACSFCS
survey \citep[see][and references therein for details on the
  observations and the data reduction procedures for the ACSFCS and
  its twin ACSVCS survey]{cote04,jordan07}\footnote{All information on
  both surveys are also available via web at the URL:
  www.astrosci.ca/users/VCSFCS}.

Other ACS observations in the $F435W$, $F555W$, and $F814W$ ($\sim BV$
and $I$, respectively) bands are also available from the HST
archive. The $B$-band data are not suitable for SBF analysis, while
the $V$ and $I$ band have already been analyzed by us
\citep{cantiello07b}, providing results in good agreement with the
present study (see below). Such observations will be used to improve
the mapping of the dust in NGC\,1316. The image processing, including
cosmic-ray rejection, alignment, and final image combination, is
performed with the APSIS ACS data reduction software \citep{blake03},
also used for ACSFCS image alignment.

For sake of homogeneity with the other datasets used in this work, our
photometric analysis of the ACS frames was performed independently
from the ACSFCS one. Hence, it is a noteworthy result that our
$g_{F475W}$ and $z_{F850LP}$ surface brightness profiles, derived as
described in the following section, match within $<0.05$ mag  the
ACSFCS ones from \citet{cote07}.  

The situation is a bit more complex for the photometry of point-like
sources and/or ``slightly-resolved'' ones. For this comparison we
used the preliminary catalog of GC candidates from the ACSFCS team,
derived as described in \citet[][]{jordan04,jordan09}, and our
photometric catalog obtained as described in Section
\ref{sec_measure}. The results of the comparison are shown in Figure
\ref{fcs}. If the full sample of $\sim$620 matching sources is used,
the median difference between our and ACSFCS photometry overlaps with
zero within the $rms$ scatter. The large scatter between the catalogs
is due to the independent analysis procedures, especially in
the way slightly-resolved sources are treated. At the distance of the
Fornax cluster with the resolution of ACS, the GCs hosted by NGC\,1316
appear slightly resolved. The ACSFCS analysis is optimized to generate
accurate photometry of GCs with different radii. The aperture
correction for such sources needs to be evaluated using more refined
analysis methods \citep[]{jordan04,jordan09} than the ones adopted
here (see next section). The gray circles in Figure \ref{fcs} show a
selection of GC candidates with $i)$ $\Delta m\leq0.05$ mag and
brighter than $m=24$ mag in both ACS bands, $ii)$ galactocentric
radius $\geq60\arcsec$, to reduce the number of objects highly
contaminated by dust, and $iii)$ ACSFCS estimated radius
$\leq0.03\arcsec$, to select only the most compact sources, for which
the issue of a different treatment of the aperture correction should
not complicate the comparison. For the selected sample of sources the
matching is significantly improved (gray symbols in the figure). The
$g_{F475W}$ data are statistically consistent in the two catalogs,
while a small 0.01 mag offset is seen in the $z_{F850LP}$
band. However, for the purposes of the present work, such an offset only
affects the estimate of the contribution to the fluctuation amplitude
due to {\it external} sources
\citep[][]{tal90,sodemann95,blake99}. Given the level of completeness
of the GC catalog, and the amplitude of the offset, the impact on
$\bar{z}_{F850LP}$ is negligible. As will be shown in section
\ref{sec_measure}, in fact, our SBF measurements are in very good
agreement with the ACSFCS ones \citep{blake09}\footnote{We adopt as
  reference the photometric VEGA zero points, while ACSFCS results use
  the AB ones. All ACSFCS data, including the calibration of SBF
  magnitudes, are transformed to the VEGA magnitude system using
  \citet{sirianni05} zero point transformations.}.

\subsubsection{NGC\,1316 as seen by the WFC3}

   \begin{figure*}
   \centering
   \includegraphics[width=0.96\textwidth]{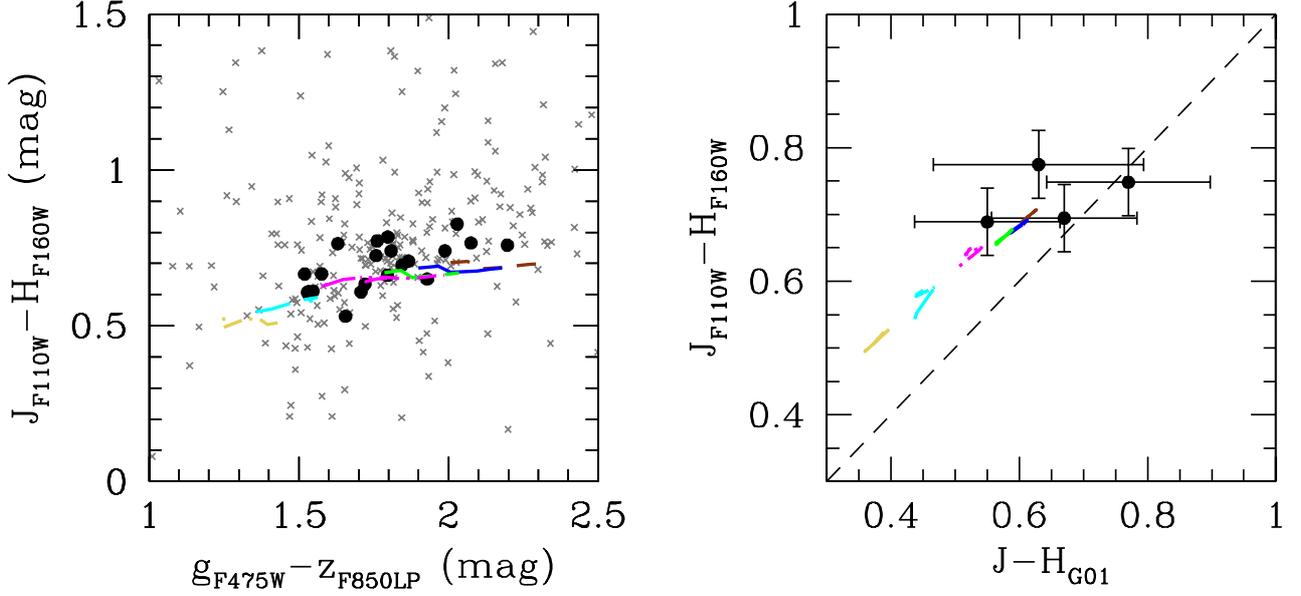}
      \caption{Left panel: colour-colour diagram with the ACS and
        WFC3/IR magnitudes. Gray crosses mark the full sample of
        common sources, black circles show selected GC candidates (see
        text). Right panel: Comparison between WFC3/IR photometry and
        the IRAC2 data from \citet{goudfrooij01a}. In both panels the
        predictions from SPoT simple stellar population models with
        different metallicity are shown with different colour/line
        styles ([Fe/H]=$-$2.3, $-$1.4, $-$0.7, $-$0.4, 0.0 and +0.4 dex shown
        with dark-yellow, cyan, magenta, green, blue and brown,
        respectively). The age range is 3 to 14 Gyr. \it{[See
            electronic version of the Journal for a colour version of
            the figure.]}}
         \label{nircal}
   \end{figure*}

The $F110W$ ($\sim J$) and $F160W$ ($\sim H$) band images of NGC\,1316
taken with the WFC3/IR were downloaded from the Hubble Legacy Archive
together with the WFC3/UVIS $F336W$ ($\sim U$-band) image. The
near-IR images were downloaded for the specific purpose of deriving
SBF magnitudes, while the $U_{F336W}$ has been used to improve the
detection and masking of dust over the entire set of available images.

We calibrated the WFC3/IR photometry using the VEGA zero points given
by \citet{kalirai09}. To check the WFC3/IR photometric data, we made
two different comparisons, both shown in Figure \ref{nircal}.
Figure \ref{nircal} (left panel) plots the WFC3/IR colour
$J_{F110W}-H_{F160W}$ versus the ACS $g_{F475W}-z_{F850LP}$ for the
combined ACS-WFC3 data set. The crosses give the full sample of
sources. In order to select the best GC candidates, the black filled
circles show the sources with $z_{F850LP}\leq 23.5$ mag, photometric
error $\Delta z_{F850LP}\lsim 0.1$ mag, SExtractor class-star
parameter $\geq$ 0.7 , and galactocentric distance $\geq
45\arcsec$. The simple stellar population (SSP) models from the Teramo
Stellar Population Tools (SPoT, see below) group\footnote{See
  www.oa-teramo.inaf.it/spot} for the age range from 3 to 14 Gyr, and
[Fe/H] from $-$2.3 dex to +0.4 dex, are also shown in the figure. Even
though a non-negligible scatter exist between the data of selected
GCs candidates and SSP models ($\lsim 0.2$ mag), the match with the
locus of SSP models is satisfactory for the purposes of the present
work.

The right panel of Figure \ref{nircal} shows the comparison between
the combined WFC3/IR and G01 near-IR samples. The SPoT models are also 
shown in the figure, and sources enshrouded in dust 
are rejected from the comparison. Notwithstanding the large error bars
of ground-based observations, we find $\Delta
[(J{-}H)_{G01}{-}(J_{F110W}{-}H_{F160W})\sim0.08]$ mag, in agreement
with model predictions.

\section{SBF measurements}
\label{sec_measure}
To derive the photometry of sources in each of the selected VLT and
HST frames, and measure the fluctuation amplitudes, we used the
procedures described in our previous works \citep[see][and references
  therein]{cantiello11a}. The procedure is basically the same for
VLT/FORS1, HST/ACS, and HST/WFC3/IR data with minor differences
outlined below.

   \begin{figure*}
   \centering \includegraphics[width=0.96\textwidth]{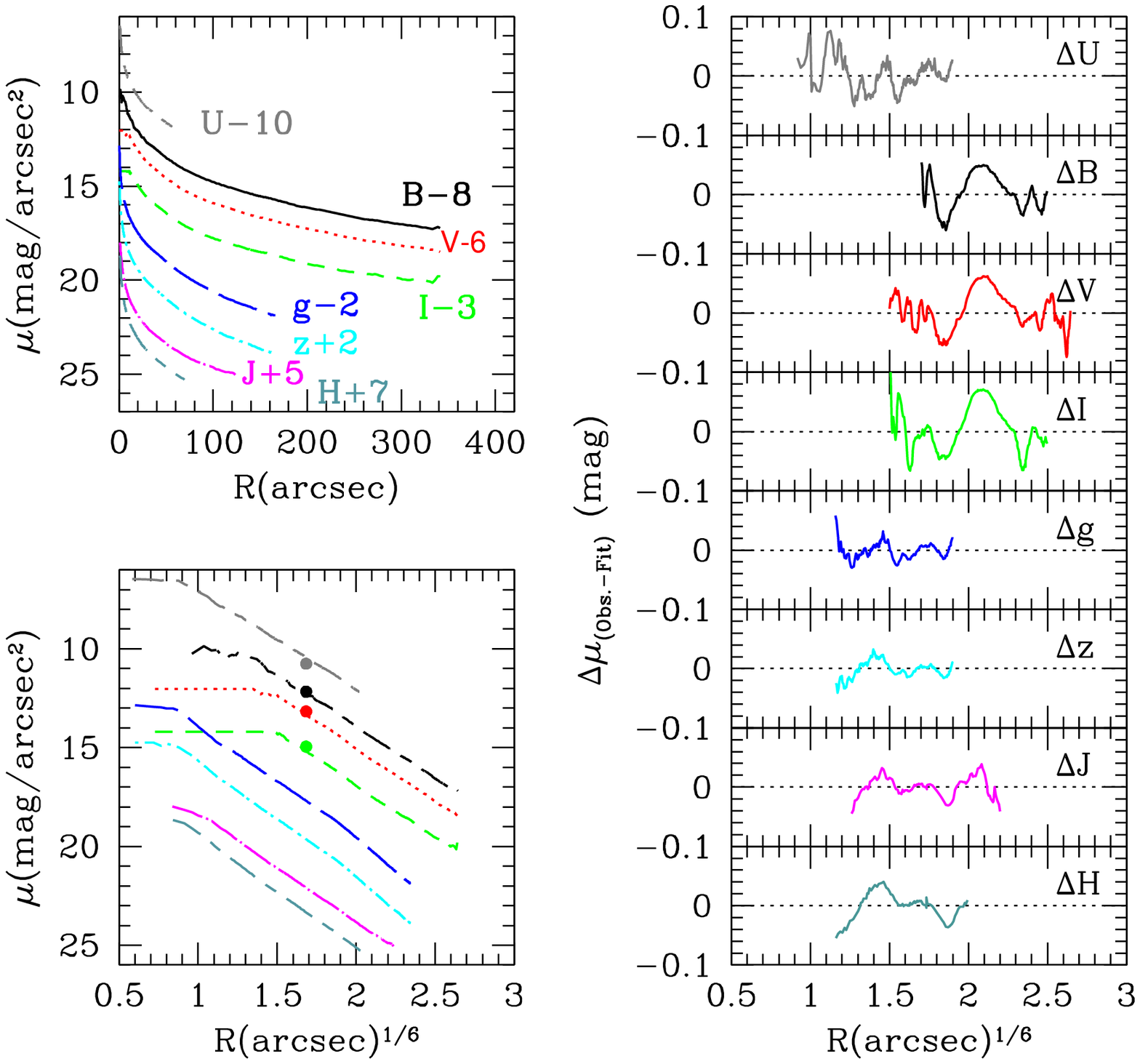}
      \caption{Upper left panel: $U_{F336w}$/WFC3/UVIS, $BVI$/FORS1,
        $g_{F475W}z_{F850LP}$/ACS, and $J_{F110W}H_{F160W}$/WFC3/IR
        surface brightness profiles for NGC\,1316.  Each band is shown
        with a different line style/colour. A vertical shift is
        applied for sake of clarity as labeled. Lower left panel: As
        upper left panel, but in $R^{1/6}$ scale. The Sersic profile
        fitting procedure used leaves the $n$-index free to
        vary. Here we use the median $n=6$ value, which is also
        consistent with \citet{cote07}. The full dots are extrapolated
        from \citet{carter83}, accordingly shifted for each band (see
        colour image). Right panels: the difference between observed
        and fitted surface brightness profiles.  \it{[See electronic
            version of the Journal for a colour version of the
            figure.]}}
         \label{muplots}
   \end{figure*}

The main steps of SBF measurement involve: sky background
determination and subtraction; galaxy model and large scale residual
subtraction; photometry and masking of point-like and extended sources, 
including dust; power spectrum analysis of the residual frame.
We determined the sky background by fitting the surface brightness
profile of the galaxy with a Sersic law \citep{sersic68} plus a
constant term. After sky determination, a first model of the galaxy
was obtained and subtracted from the sky-subtracted frame, and a mask
of the bright sources was obtained. The large scale residuals, still
present in the frame after subtracting the galaxy model, were removed
using the background map obtained with SExtractor adopting a mesh size
$\sim 10$ times the FWHM \citep{tal90,cantiello05}. In the following
we refer to the sky, galaxy-model and large scale residuals subtracted
image as the {\it residual} frame.

The procedure of $i)$ surface brightness analysis and sky
determination, $ii)$ model fitting and subtraction, $iii)$
sources/dust masking, and $iv)$ large scale residual subtraction was
iteratively repeated until the residual frame appeared ``flat'' in the
regions of interest for SBF measurements, i.e., until the residual did
not show any (local) artifact due to the subtracted galaxy model. The
middle panel of Figure \ref{img_res} show the false colour combined
$BVI$ image of the residual frames. The right panel of the figure,
instead, shows a combination of the ACS and WFC3/UVIS frames, used to
map the dust around the center of the galaxy. The positions of three
of the four SNe~Ia host in the galaxy is also indicated in the figure
\citep[the region of SN\,1980D is not covered by either the HST or VLT
  frames; we adopted the revised SNe Ia coordinates
  from][]{stritzinger10}.

The surface brightness profiles for all bands analyzed, as well as the
difference between modeled and observed profiles, are shown in Figure
\ref{muplots}. 

The photometry of fore/background sources and of GCs was derived
running SExtractor on the residual frames. As described in our
previous work, we modified the input weighting image of SExtractor by
adding the galaxy model \citep[times a factor between 0.5 and 10,
  depending on the expected amplitude of the SBF signal; for details
  see][]{jordan04,cantiello05} so that the SBFs were not detected as
real objects. The aperture correction was obtained from a number of
isolated point-source candidates in the frames and by making a curve
of growth analysis out to large radii
\citep[][]{cantiello09,cantiello11b}. The outer radius used for FORS1
data was $6\farcs0$. For ACS(WFC3) we adopted
  $0\farcs8$($1\farcs6$) and then added an extra aperture correction
term to {\it infinite} radius by using the instrument encircled energy
tables \citep[for ACS we used][while for WFC3 we adopted the
  Instrument Handbook, version 4]{sirianni05}

Once the catalog of sources was derived, the next step was to fit the
luminosity function of the sources, to be used to estimate the
already mentioned background fluctuation term due to unmasked faint
sources. We obtained the fit to the GC and background galaxy luminosity
functions from the photometric catalog of sources, after removing the
brightest/saturated point-like sources and the brightest and most
extended objects. The best fit to the sum of the two luminosity
functions, and the background fluctuation correction term, $P_r$, were
derived as in \citet{cantiello05}.

   \begin{figure*}
   \centering
   \includegraphics[width=0.96\textwidth]{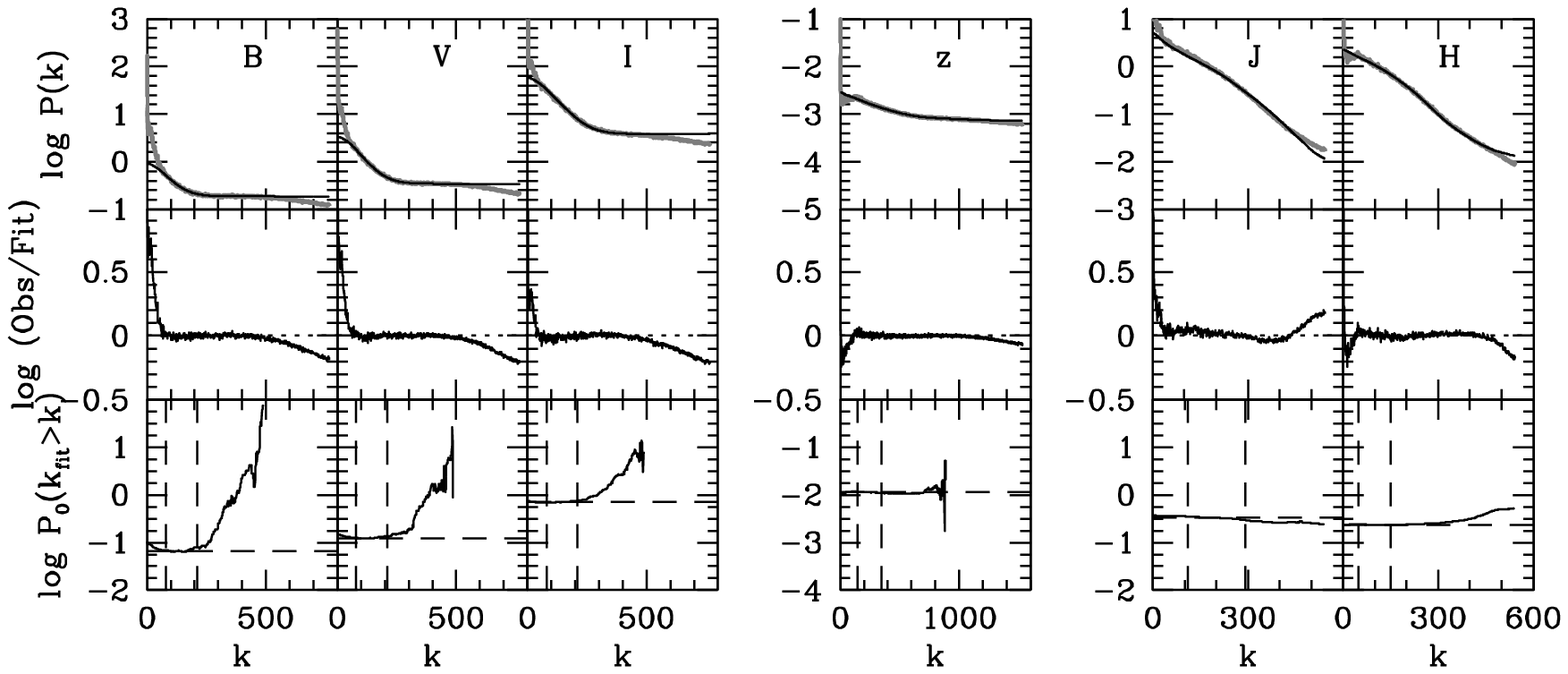}
      \caption{Power spectrum analysis of the
        $BVIz_{F850LP}J_{F110W}H_{F160W}$ frames. Each panel shows a
        different band, as labeled. For all bands, the upper panels
        show the logarithm of the power spectrum of the residual frame
        (gray dots) and the best fit $P(k)$ curve. In the middle
        panels the difference between observed and fitted power
        spectra is shown. The flat region of $\log~P_0(k_{fit}>k)$
        between vertical dashed lines (lower panels) is used to
        evaluate the best fit parameters $P_0$, and $P_1$.}
         \label{pofit}
   \end{figure*}

To measure SBF magnitudes we estimated the azimuthal average of
the residual frame power spectrum, $P(k)$, then matched it to
the power spectrum of a template PSF convolved with the mask image,
$E(k)$. The total fluctuation amplitude $P_0$ was obtained via a
robust minimization method \citep{press92} as the multiplicative
factor in the power spectrum representation $P(k)=P_0 \times E(k) + P_1$,
where $P_1$ is the white noise constant term. We used one to five
different isolated bright point sources in each residual frame for the template
PSFs.  Each PSF, after normalization, was singularly 
adopted to estimate the SBF signal of the galaxy.  Finally, the
SBF amplitude, $P_f=P_0-P_r$, was estimated within circular
annuli. The results of the power spectrum analysis are summarized in
Figure~\ref{pofit}, with one example for each band.

The results of the SBF and colour measurements for all bands considered
are reported in Table \ref{tab_sbf}. For each filter, in addition to
the statistical error, we report the systematic uncertainty due to PSF
fitting.  For $H_{F160W}$, since we could only find one good candidate
PSF in the frame, we assumed a conservative PSF scatter of 0.2
mag\footnote{Previous studies indicate SBF variations $<0.1$ mag with
  PSF \citep[e.g.,][]{liu02,cantiello05,blake10b,cantiello11b}.}. In
the case of $J_{F110W}$ no good PSF was found over the WFC3/IR frame,
thus we used a PSF star taken from different observations associated
with the same HST proposal. As for $H_{F160W}$ we assumed 0.2 mag PSF
uncertainty.

\begin{table*}
\caption{Surface Brightness Fluctuation and colour measurements corrected for galactic extinction.}
\label{tab_sbf}
\centering
\begin{tabular}{c c c c c c}
\hline\hline 
\multicolumn{6}{c}{VLT data} \\
\hline 
$\langle r \rangle$ &  $B{-}V$ & $V{-}I$ &  $\bar{B}$ &  $\bar{V}$ &$\bar{I}$  \\
    (arcsec)        &    (mag) &   (mag)  & (mag)     &  (mag)     &  (mag)    \\
 170    &   0.84 $\pm$0.03 &   1.116 $\pm$0.015 &  33.02 $\pm$0.02 &  32.19 $\pm$0.09 &  29.81 $\pm$0.07  \\
\hline
\multicolumn{3}{c}{PSF uncertainty} & 0.09 & 0.04 & 0.03 \\
\hline\hline 
\multicolumn{6}{c}{ACS data} \\
\hline 
 $\langle r \rangle$ &  $g{-}z$ &         &  &          &   $\bar{z}$       \\
     (arcsec)        &    (mag) &         &  &          & (mag)             \\
 80                  &  1.954 $\pm$0.011 &     &  &          & 29.11$\pm$  0.06        \\
\hline 
\multicolumn{3}{c}{PSF uncertainty} &    &  & 0.014 \\
\hline\hline 
\multicolumn{6}{c}{WFC3/IR data} \\
\hline 
$\langle r \rangle$ &  $V{-}I$$^{\mathrm{a}}$ & $J{-}H$        &  &   $\bar{J}$       &   $\bar{H}$       \\
    (arcsec)        &    (mag) &                &  &   (mag)           & (mag)             \\
 80  & 1.102 $\pm$0.002 & 0.72 $\pm$0.01 & &   27.32$\pm$ 0.07&   26.28 $\pm$0.04  \\
\hline 
\multicolumn{3}{c}{PSF uncertainty} &    &  0.2 & 0.2 \\
\hline\hline 
\end{tabular}
\begin{list}{}{}
\item
[[$^{\mathrm{a}}$]] Colour data obtained from VLT data using the same
masks adopted for WFC3/IR measurements.
\end{list}
\end{table*}


\section{The SBF distance to NGC\,1316}
\label{sec_distances}

\begin{table*}
\caption{Distances from SBF  measurements.}
\label{tab_distances}
   \begin{center}
\begin{tabular}{l r c c}
\hline\hline 
Passband          & $\bar{M}$ calibration eq.                   &  $(m-M)$       & Reference        \\
\hline
\multicolumn{4}{c}{Empirical calibrations} \\
\hline 
$V$               & $(0.83\pm0.12)+(5.3\pm0.8)[(V{-}I)-1.15]$   & 31.53$\pm$0.17 &  [1]             \\ 
$I$               & $(-1.68\pm0.08)+(4.5\pm0.25)[(V{-}I)-1.15]$ & 31.65$\pm$0.12 &  [2]             \\ 
$I$               & $(-1.6\pm0.1)+(3.0\pm0.3)[(B{-}I)-2.0]$     & 31.55$\pm$0.13 &  [3]             \\ 
$z$  & $-$2.04+1.41$x$+2.60$x^2$+3.72$x^3$, $x\equiv(g{-}z)-1.94$& 31.66$\pm$0.07 & [4]  \\
$J_{F110W}$               & \nodata                                     & \nodata       & \nodata                      \\
$H_{F160W}$               & $(-4.8\pm0.1)+(5.1\pm0.5)[(V{-}I)-1.16]$    & 31.3$\pm$0.2  &   [5]        \\ 
$H_{F160W}$               & $-5.17+0.70x+2.90x^2$, $x\equiv(g{-}z)-1.94$      & 31.4$\pm$0.2   &   [6]        \\ 
\multicolumn{2}{r}{Weighted Mean}                               &31.59$\pm$0.05 &                              \\
\hline\hline 
\multicolumn{4}{c}{Theoretical calibrations} \\
\hline 
$V$ &   $0.89+4.01 [(V{-}I)-1.15]$  (0.3)                          & 31.4 $\pm$0.3   &                     \\
$I$ &  $-1.63+5.11 [(V{-}I)-1.15]$  (0.3)                          & 31.6 $\pm$0.3   &                     \\
$I$ &  $-1.63+2.65 [(B{-}I)-2.0] $  (0.3)                          & 31.6 $\pm$0.3   &                     \\
$z$ &  $-2.77+2.06 [(g{-}z)-1.94]$  (0.2)                         & 31.8 $\pm$0.2 &                     \\
$J_{F110W}$ &  $-3.75+3.07 [(V{-}I)-1.16]$  (0.3)                          & 31.3 $\pm$0.3   &                     \\
$H_{F160W}$ &  $-4.86+3.59 [(V{-}I)-1.16]$  (0.3)                          & 31.4 $\pm$0.4   &                     \\
\multicolumn{2}{r}{Weighted Mean}                                  & 31.60$\pm$0.11  &                        \\
\hline 
\end{tabular}
	\end{center}
\tablefoottext{1}{\citet{bva01}}\\ \tablefoottext{2}{\citet{tonry01}
  with revised Cepeheid
  distances}\\ \tablefoottext{3}{\citet{cantiello05}}\\ \tablefoottext{4}{Calibration
  using \citet{blake09}, uncertainties evaluated from eq. (1) in
  \citet{mei07xiii}}\\ \tablefoottext{5}{Calibration from
  \citet{jensen03} with metallicity correction on Cepheids and NICMOS to
  WFC3 $H_{F160W}$ zeropoint correction (see text).}\\ \tablefoottext{6}{Calibration from
  \citet{cho13}.}\\
\end{table*}


The estimate of distances in the SBF method
relies on knowledge of the absolute SBF magnitudes. Using the
measurements reported in Table \ref{tab_sbf}, together with empirical
and theoretical calibrations given in Table \ref{tab_distances}, we
obtained the distance moduli reported in the $(m-M)$ column of the
table. In particular we obtained a mean distance modulus of
31.59$\pm$0.05 mag and 31.60$\pm$0.11 mag with the empirical and
theoretical equations, respectively.
In the following we provide some details on the calibration equations
adopted.

\subsection{Absolute SBF magnitudes from empirical calibrations}
\label{sec_empirical}
The empirical calibrations of SBF magnitudes in optical bands, in
particular in the $I$ band, are the most thoroughly analyzed
\citep{tal90,tonry01,mei07xiii,blake10b}.  The two
aforementioned HST surveys of the Virgo and Fornax clusters provided
an extremely accurate calibration of $\bar{z}_{F850LP}$, including
some degree of non-linearity in the calibration.
Some debate still exists on near-IR bands calibrations
\citep[see][and references therein]{gonzalez10}, although, as already
mentioned, relevant progress will be done thanks to the installation
of the WFC3/IR \citep[][]{blake12,french12}.

In the upper part of Table \ref{tab_distances} we report the distance
moduli obtained using the empirical calibrations taken from
literature, together with the adopted calibrations.

As a first general comment on the empirical equations, {\it we must
  emphasize that the numbers reported in the table are all tied to the
  same common zero point, i.e., to the Cepheid distances with
  metallicity correction to the PL relation from \citet{freedman01}.}
The resulting zero points of the $\bar{V}$, $\bar{I}$ and
$\bar{H}_{F160W}$ versus $V{-}I$ calibrations are shifted of $+0.06$,
$+0.06$ and $-0.10$ mag with respect to the calibrations in the
original papers \citep[see appendix A in][]{blake10b}.

For the $\bar{I}$ versus $B{-}I$ calibration, taken from
\citet{cantiello05}, we do not make any revision since the zero point
is already based on the chosen set of Cepheids. Similarly, the
$z_{F850LP}$-band calibration does not need any change
\citep{blake09}.

For the $H_{F160W}$-band distance estimate, we used two independent
calibrations.  The first derived by \citet{jensen03} from HST/NICMOS
data. We adopted the SPoT SSP models to evaluate the changes in the
calibration due to the difference between the NICMOS2 and the WFC3/IR
$H_{F160W}$ passbands. The result is that the WFC3/IR
$\bar{H}_{F160W}$ zero point is 0.2 mag fainter than the NICMOS2
$H_{F160W}$ one. This is partly expected because of the cut at
larger/redder wavelengths of the WFC3/IR filter (1400-1700~nm
passband, versus 1400-1800 nm for NICMOS2). Hence, we add a 0.2 mag
to the zero point of the \citet{jensen03} empirical relation, assuming a
default 0.1 mag uncertainty because of the model-dependent
correction term.
The second calibration is a preliminary result obtained by
\citet{cho13}, based on the observations of 16 early-type galaxies in
Virgo and Fornax specifically obtained to empirically calibrate the
SBF for the WFC3/IR passband.
For the near-IR data, in contrast with the optical measurements, the
distance modulus includes the systematic PSF uncertainty since it is
dominant with respect to the statistical errors of the SBF measurement
and of the calibration zero point.
 
All distances based on the empirical calibrations reported in Table
\ref{tab_distances} agree with each other within the quoted
uncertainties. The weighted mean of distance moduli is also given in
the table.

\subsection{Absolute SBF magnitudes from theoretical calibrations}
\label{sec_theory}

Various authors have analyzed the possibility to calibrate absolute
SBF magnitudes using stellar populations synthesis models, thereby
making it a primary distance indicator, not linked to the Cepheids
zero point \citep[][]{buzzoni93,worthey93a,bva01,biscardi08}.  In
this work we have taken as reference the SBF versus colour equations
derived using the simple stellar population models from the Teramo
SPoT group. For a detailed review of the SPoT models we refer to
\citet{raimondo05} and \citet{raimondo09}, and references therein.
These models have been shown to be very effective in matching the
empirical SBF calibration in different bands, as well as in
reproducing the resolved (colour magnitude diagrams) and unresolved
(colours, magnitudes) properties of stellar populations
\citep{brocato00,cantiello07b,cantiello12}. We used the updated
version of the SPoT models, which for the photometric bands and
chemical composition used in this section confirms the results
obtained from the previous \citet{raimondo05} models (G.~Raimondo,
private communication). The grid of models used has [Fe/H]= $-$0.4,
0.0,$+$0.4 dex and ages from 3 to 14 Gyr. The choice was made based on
the age and chemical composition properties of fields stars in
NGC\,1316 derived from various independent spectro-photometric
indicators \citep[e.g.,][]{terlevich02,silva08,konami10}, and also
confirmed by GC analysis \citep[][see also \S
  \ref{sec_ssp}]{goudfrooij01a,goudfrooij01b}. Under these
assumptions, and using a bootstrap approach, we obtained the
calibration equations reported in the lower part of Table
\ref{tab_distances}. For each equation, the scatter of the $\bar{M}$
versus colour relation is also tabulated. The uncertainties on
distance moduli are derived by summing in quadrature the scatter of the
theoretical calibration and the uncertainty on $\bar{m}$.

The distance moduli derived with the theoretical calibrations from
the SPoT models are given in the Table \ref{tab_distances}. All reported
distances agree to within the quoted uncertainties.

\subsection{Combining SBF-based distance moduli (plus a fundamental note on uncertainties)}

The SBF distance of NGC\,1316 presented in \S \ref{sec_empirical} and
\S \ref{sec_theory} is based on a self-consistent treatment of
empirical and theoretical SBF calibrations. Empirical calibrations
have been tied to the same common zero point reference. On the other
hand, the theoretical calibrations are based on a well defined set of
SSP models ranging from $V$ to $H_{F160W}$, i.e., a wavelength interval
that, in terms of SBF and colours, samples very different stellar
population (sub)components.

It is noteworthy that the {\it empirical} and {\it theoretical}
evaluations agree very well with each other. Taking into account that
the two methods are based on independent calibration
procedures, subject to different types of systematic and statistical
uncertainties, this result suggests that both types of errors are
reasonably well constrained.

For what concerns the systematic errors in our measurements, summing
up all expected sources of uncertainty for each one of the three
instruments considered -- filter zero point, data reduction,
calibration zero point, PSF normalization -- the expected systematic
uncertainty is $\lsim$0.1 mag, with the exception of near-IR bands,
where the contribution from the PSF normalization is dominant with
respect to the others.

For the systematic errors in the empirical calibrations, recall that
all relations are linked to the same Cepheid zero point
\citep{freedman01}, which accounts for a further systematic $\lsim0.2$
mag uncertainty\footnote{A further component to the systematic error
  comes from bandpass mismatch with various telescopes. \citet{bva01}
  presented a discussion of this issue (see their Sect. 5.5, Fig. 15
  in particular), showing that at the colour of the SBF,
  $(\bar{V}{-}\bar{I})\sim2.4$ for NGC\,1316, the difference in the
  standard Cousins $I$ and the HST $I$ can vary by $\pm0.02$ mag.}.

   \begin{figure}
   \centering
   \includegraphics[width=0.48\textwidth]{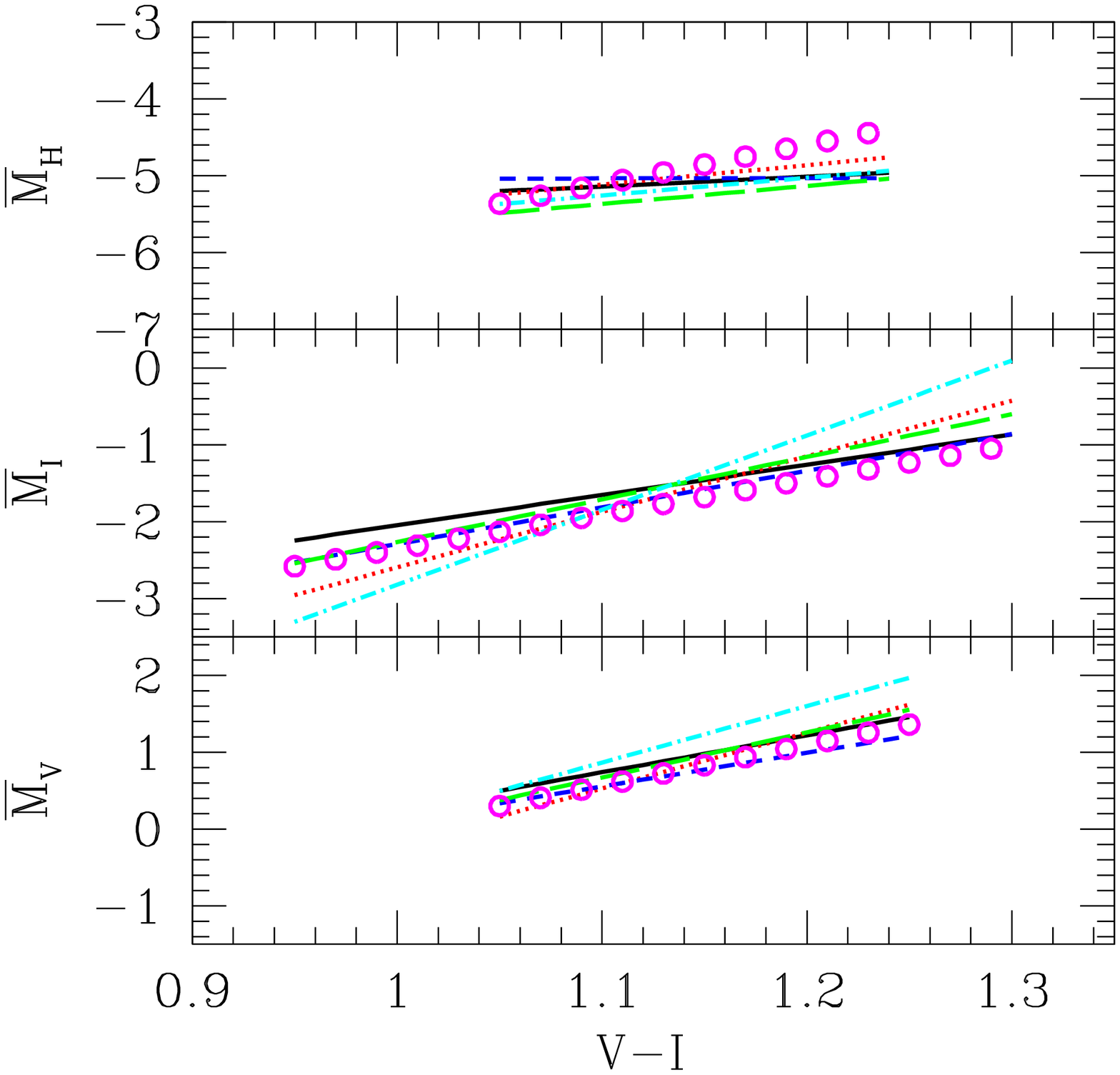}
      \caption{$V$, $I$ and $H_{F160W}$-band SBF versus $(V{-}I)$
        predictions from five independent stellar population
          synthesis codes: solid line (black), dotted line (red),
          dashed line (blue), long-dashed line (green), dot-dashed
          (cyan) line show the updated SPoT, \citet{bva01},
          \citet{liu02}, \citet{marin06}, and \citet{gonzalez10} best
          fit lines, respectively. Empty (magenta) circles show the
          corresponding empirical relations, also given in Table
          \ref{tab_distances}.  \it{[See electronic version of the
            Journal for a colour version of the figure.]}}
         \label{modscatter}
   \end{figure}

Estimating the systematic uncertainty in theoretical SBF calibrations
is not a simple task. One way to get such an estimate would be to
change the ingredients in the SSP code (stellar tracks, initial mass
function, atmosphere models, etc.), and then analyze the effects on
SBF. A rough (indirect) estimation of the uncertainty on theoretical
calibrations is to compare the results from independent models,
obtained from different SSP codes relying on independent physical
input and algorithms. A first attempt along these lines was carried
out by \citet{cantiello03}. To estimate this uncertainty for the
present work, we compare SSP model predictions from \citet{bva01},
\citet{liu02}, \citet{marin06}, \citet{gonzalez10} and the SPoT
models. For all these models, we obtain linear fits to the $V$, $I$
and $H_{F160W}$ SBF amplitudes versus $(V{-}I)$ colour (Figure
\ref{modscatter}). The empirical calibration for each band is also
shown in the figure. Note the different range of colour used for each
equation, depending on the range of validity of the empirical equation
\citep{tonry01,bva01,jensen03,cantiello07b}.  We find that a scatter
of $\sim$0.2 mag provides a first approximation to the systematic
uncertainty of SBF calibrations derived from SSP models. A discussion
on the origin of the scatter between models is beyond the scope of
this paper. We refer the interested reader to the quoted papers and
references thereafter.

In conclusion, by combining the weighted mean distance moduli,
$(\bar{m}-\bar{M})_{empirical}=31.59\pm0.05({\rm stat.})  \pm0.20
({\rm sys.})$ mag and $(\bar{m}-\bar{M})_{theoretical}=31.60\pm0.11 ({\rm
  stat.})\pm0.20 ({\rm sys.})$ mag, we conclude that our best estimate
of the distance of NGC\,1316 is $(\bar{m}-\bar{M})=31.59\pm0.05({\rm
  stat.})\pm0.14({\rm sys.})$ mag, or $d=20.8\pm0.5({\rm
  stat.})\pm1.5({\rm sys.})$ Mpc.

\section{Comparison with distances from the literature}
\label{sec_compare}

As already mentioned, the VLT observations used here were obtained to
detect and study the population of novae in NGC\,1316, and to use them
to derive the galaxy distance. \citet{dellavalle02}, using the data of
four detected novae (the first detected beyond the Virgo cluster at
the time), and the Buscombe-de Vaucouleurs relation
\citep{buscombe55,capaccioli89,capaccioli90} set an upper limit on the
distance of the galaxy equal to $22.4$ Mpc, and a lower limit
of $16$ Mpc from the data of Nova~A in their sample. In
spite of the difficulties (observational and statistical) in using the
novae in Fornax A to estimate the galaxy distance, the range obtained
by \citeauthor{dellavalle02} is in excellent accord with our
estimates. Moreover, the lower limit distance is also in complete
accord with the SNe Ia as well as the PNLF distances (see below).

In the following sections we compare our distance estimates with
others available in the literature. We consider the results obtained
from three different distance indicators: Type Ia supernovae, the
planetary nebula luminosity function (PNLF), and mean properties of
the GC system. Other distances indicators, such as the Tully-Fisher
relation or the fundamental plane, cannot be used reliably for
NGC\,1316 because of its irregular post-merger morphology; moreover,
these indicators are normally applied to groups or clusters as a whole,
rather than giving precise distances to individual galaxies.

\subsection{Comparison with SNe~Ia}

\citet{ajhar01}, and later \citet{cantiello11a}, have presented a
comparison of SBF and Type Ia SNe distances for a total of 15 different
SNe~Ia in 14 galaxies. Both studies found excellent overall agreement between
the two distance indicators, including the case of NGC\,1316, provided
that a consistent set of Cepheid-based distances is used.  
Similarly, \citet{freedman01} and \citet{freedman10} derive
essentially identical values of $H_0$ from these two methods when
calibrated consistently via Cepheids.

\begin{table*}
\caption{Type Ia Supernova distances to NGC\,1316 available in the literature.}
\label{tab_sne}
\centering
\begin{tabular}{l c c c}
\hline\hline 
SN      & $\mM$            & Method   &  References/Notes         \\
\hline 
\multicolumn{4}{c}{Measurements from Str10} \\
80N+81D+06dd & 31.180 $\pm$ 0.013 & SNooPy EBV & Author's best estimate           \\
80N+81D+06dd & 31.25 $\pm$ 0.03 & Tripp      & Author's best estimate              \\
80N+81D+06dd & 31.203 $\pm$ 0.012 & $(m-M)_{max}$ Near-IR   & Author's best estimate         \\
SN\,2006mr     & 31.83 $\pm$ 0.07   & Tripp    & Flagged as doubtful        \\
SN\,2006mr     & 31.739 $\pm$ 0.005 &$(m-M)_{max}$ Near-IR    &     Flagged as doubtful     \\
\hline\hline 
\multicolumn{4}{c}{Other Measurements$^{\mathrm{a}}$} \\
SN\,1980N & 31.45             & $M_{max}$& \citet{hamuy91}\\
SN\,1981D & 31.35             & $M_{max}$& \citet{hamuy91}\\
\hline 
SN\,1980N & 31.44 $\pm$0.32   & MLCS     & \citet{krisciunas00}\\
SN\,1981D & 31.03 $\pm$0.3    & MLCS     & \citet{krisciunas00}\\
\hline 
SN\,1980N & 31.30 $\pm$0.15 & MLCS  & \citet{ajhar01}      \\
SN\,1980N & 31.51 $\pm$0.10 & $\Delta m_{15}$  & \citet{ajhar01}      \\
\hline 
SN\,1980N & 31.35             & $M_{max}$& \citet{reindl05}\\
SN\,1981D & 30.98             & $M_{max}$& \citet{reindl05}\\
\hline 
SN\,1980N & 31.38 $\pm$0.13 & MLCS2k2  & \citet{jha07}      \\
SN\,1981D & 30.78 $\pm$0.24 & MLCS2k2  & \citet{jha07}      \\
\hline 
SN\,2006dd & 31.34 $\pm$0.11 & MLCS   & \citet{maoz08}      \\
\hline\hline 
\end{tabular} \\
\tablefoottext{a}{All distances  are converted to a scale with
  $H_0=72~km~s^{-1}~Mpc^{-1}$.}
\end{table*}

NGC\,1316 has been a prolific producer of Type Ia supernovae, with
four recorded events: SN\,1980N, SN\,1981D, SN\,2006dd, and the fast
declining SN\,2006mr. All SN\,Ia light-curves have been used to
determine the distance of the host galaxy using several
methods. Furthermore, being one of the nearest bright galaxies with
well sampled SNe~Ia light-curves, NGC\,1316 is frequently used in
calibrating samples for high-redshift SNe~Ia
\citep[e.g.,][]{jha07,burns11}.
Based on the MLCS SN\,Ia distance method \citep{riess98},
\citet{goudfrooij01a} reported $\mM = 31.80\pm0.05$ mag for NGC\,1316
and concluded that it was therefore $\sim0.25$ mag more distant than
the rest of the Fornax cluster, for which \citet{ferrarese00} gave a
mean distance modulus of 31.54~mag from Cepheids and other indicators.
However, no details on the calibration, etc., were given by
\citet{goudfrooij01a}. \citet{ajhar01} reported both MLCS and
$\Delta{m}_{15}$ distances \citep{phillips93,hamuy96} for SN\,1980N in
NGC\,1316 under different calibrations. Rescaling their results to
$H_0=72~km~s^{-1}~Mpc^{-1}$ gives $\mM=31.30\pm0.15$ and $\mM=31.51\pm0.10$ from the MLCS
and $\Delta{m}_{15}$ methods, respectively. Interestingly, the
$\Delta{m}_{15}$ value coincides exactly with the mean SBF distance
for the Fornax cluster by \citet{blake09}.

On the other hand, \citet[][Str10 hereafter]{stritzinger10} have
recently reanalyzed the SN\,Ia distance to NGC\,1316, and obtained
$(m-M)=31.25 \pm 0.03$ ({\rm stat.}) $\pm 0.04$ ({\rm sys.}) mag,
based on the analysis of the three ``normal'' SNe~Ia (adopting
$H_0=72~km~s^{-1}~Mpc^{-1}$, see Table \ref{tab_sne}). The authors
also obtain values of $(m-M)\sim 31.7-31.8$ mag from the data of the
fast declining SN\,2006mr; however, they consider as doubtful the
suitability of fast-declining SNe~Ia for estimating distances. Thus,
the best distance modulus to NGC\,1316 obtained by Str10 is $\sim$0.34
mag fainter than ours, equivalent to a 17\% smaller distance. Note that 
the new distances derived by Str10, as pointed out by the authors, are
based on a thorough analysis and discussion of the four SNe Ia, while
little or no detail were given in previous studies. Hence, such a
difference is very interesting, and needs further analysis, especially
because of the primary role in the cosmological distance scale of both
these indicators.

As a first general comment, it is useful to highlight that two of the
normal SNe (SN\,1980N and SN\,1981D) were observed in the pre-CCD
era. So, although they afford a nice opportunity to test the internal
consistency of SNe Ia distances, one should not forget that the
quality of the data of these objects is lower than what is currently
being obtained. Bearing this in mind, we proceed to our analysis taking
the SNe~Ia distances, and their associated systematic/statistical
errors, from the cited references.

The SNe~Ia distances to NGC\,1316 obtained before the study of Str10,
are listed in Table \ref{tab_sne}.  For consistency with our and
Str10's measurements, all distances are converted to a scale with
$H_0=72$~km~s$^{-1}$~Mpc$^{-1}$. In the table we do not report the
systematic errors, which are not given by all authors. Taking the
weighted mean of all measurements in Table \ref{tab_sne}, except those
of Str10, adopting 0.4 mag default uncertainty where no error is
reported, we obtain $(m-M)=31.35\pm0.05$ mag. Although such value
agrees with our estimates better than the Str10 result, we caution the
reader against these ``general'' averages, reported here only to
emphasize the relative distribution of distances with respect to a
reference point. Some of the values reported in Table \ref{tab_sne},
in fact, are obtained using the same methods/objects but under
different assumptions (e.g. MLCS and MLCS2k2), and/or with different
calibrators, so that the reported average does not necessarily
have a correct physical meaning.

Str10 derives the distance to NGC\,1316 using three different methods:
the EBV, the Tripp method, and the near-IR light-curves of Type Ia
SNe. Each method has its own calibration, and is quantitatively
independent from the others. Hence, there can be various possible
causes for the difference between our and Str10 distances. First, for
the EBV method, the authors adopt negligible internal extinction for
all four SNe~Ia, but also comment that the spectroscopic analysis
provides results ``totally inconsistent with the low host galaxy
reddening'' (Str10, Section 4).  However, the optical and
optical/near-IR colours of these objects are consistent
with minimal to unreddened supernova (Figure~13 of Str10). 
Second, for the near-IR method, which is intrinsically much less
affected by the host internal extinction, we find that the
calibration used, from \citet{krisciunas09}, is based in part on a
compilation of Cepheid and/or SBF distances that is not internally
consistent (see the discussion in Appendix~\ref{appendb}).

Interestingly, Str10 obtain a SN~Ia distance modulus for the Fornax
cluster member \object{NGC\,1380} of $\mM = 31.611\pm0.008$ mag, which
is consistent with the SBF result for the same galaxy of
$31.632\pm0.075$ mag \citep{blake09}, and very similar to the SBF
distance for NGC\,1316.  In fact, there is little significant
variation in the SBF distances among the magnitude-limited sample of
43 early-type Fornax galaxies studied by \citet{blake09}, and most are
consistent with the mean Fornax SBF modulus of $31.51\pm0.03$
mag\footnote{The recent Type Ia supernova SN\,2012fr
  \citep{childress12} occurred in \object{NGC\,1365}, a giant barred
  spiral galaxy in the direction of the Fornax cluster with a measured
  Cepheid distance \citep{silbermann99,freedman01}.  However, from a
  comparison of Cepheid and SNe~Ia distances, \citet{suntzeff99}
  suggested that NGC\,1365 is actually $\sim0.3$ mag in front of the
  Fornax cluster ellipticals; \citet{kelson00} came to a similar
  conclusion based on the fundamental plane.}.

Finally, we must note that the \citeauthor{tripp98}, the maximum near-IR
magnitudes, and the MLCS2k2 methods applied to the fast declining
SN\,2006mr give distance estimates in agreement with the ones
presented here. In spite of this, it should also be emphasized that
the debate on whether fast declining SNe Ia can be used to determine
precise distances is still open \citep[so that different authors
  include or not these objects in their final samples;
  e.g.][Str10]{jha07,folatelli10,burns11}.

\subsection{Distances from Globular Cluster System properties}

Distances to NGC\,1316 derived from the properties of the globular
cluster system have generally been flagged as unreliable by the
authors due to the peculiar properties of the galaxy and its GC
system.

\citet{gomez01} made one of the first attempts to constrain the
distance to NGC\,1316 with the Globular Cluster Luminosity Function
\citep[GCLF,][]{harris01}, using ESO/EFOSC2 data. The $BVI$ weighted
average distance modulus they provided is $(m-M)\sim 31.4$ mag, based
on a calibration consistent with ours. However, the authors
specifically commented on the existence of red and blue
sub-populations of GCs and concluded that their catalog was not
sufficiently deep to reliably constrain the galaxy distance.
In an earlier study of five bright Fornax cluster galaxies,
\citet{bt96} had noted that overall the GCLF of NGC\,1316 was not well fitted
by a Gaussian (reduced $\chi_n>3$, as compared to $\chi_n\approx1$ 
for the four others), and thus could not provide a reliable GCLF distance for it.
 \citet{goudfrooij04} presented a deeper analysis of the GC
system based on ACS data. These authors confirmed the previous results of
a blue GC sub-population, consistent with a Gaussian luminosity
function, and a further component of red GCs with a power law
luminosity function. More recently, \citet{villegas10} for the ACSFCS
survey, write {\it ``NGC\,1316 is [...] not included in the fits
  because the observed GC system in this galaxy is highly influenced
  by its interaction and proximity with its satellite galaxies, and
  therefore our GCLF fit is not reliable.''}

\citet{masters10} use the GC-radii method \citep{jordan05} to analyze
the distances of Fornax Cluster galaxies.  However, the authors warn
that, while the method is effective for typical GC systems, it cannot
be applied to NGC\,1316 due to the large number of extended GCs.

In conclusion, the GC--based distances reported, for example in the
NED archive, $(m-M)_{GCLF,g}=33.55$ mag, $(m-M)_{GCLF,z}=33.68$ mag,
and $(m-M)_{GC-radius}=30.59\pm0.11$ mag, are excluded from the
comparison.

\subsection{Planetary Nebulae Luminosity Functions}
\label{sec_pnlf}
\citet{feldmeier07} reported a PNLF distance modulus to NGC\,1316 of
$(m-M)=31.26^{+0.09}_{-0.12}$ mag. This result relies on the PNLF
calibration by \citet{ciardullo02}, whose zero point is tied to the
\citet{freedman01} sample of Cepheid's distance moduli {\it with no
  dependence on metallicity of the PL relations.}  As discussed above
(\S \ref{sec_empirical}), the SBF calibrations presented here are
based on the metallicity-corrected Cepheid PL relations. Hence, to
properly compare the PNLF and SBF distance, a correction is
necessary. In a recent review, \citet{ciardullo12} finds that if one
adopts the metallicity-dependent Cepeheid PL relations, the PNLF zero
point brightens by 0.07 mag, going from $M^*=-4.46\pm0.05$ to
$M^*=-4.53\pm0.04$ mag (external scatter $\sigma=0.16$ mag).
\citet{feldmeier07} actually used $M^*=-4.47$; thus their PNLF
distance modulus for NGC\,1316 becomes $(m{-}M)=31.32^{+0.09}_{-0.12}$
mag\footnote{We have not taken into account a further $-$0.04 mag
  correction term to SBF distance moduli cited by \citet[][their
    section 6]{ciardullo02}. Including such an extra correction to our
  distance modulus implies $(\bar{m}{-}\bar{M})\sim31.55$ mag.}.

Both PNLF distances given above agree within quoted uncertainties with
the Str10 distance. However, the SBF and the updated PNLF
  distances are consistent within the given statistical and systematic
  errors, notwithstanding that the difference between them remains
  non-negligible. In other words, within the given uncertainties, the
  SBF distance is consistent with the PNLF, and the latter with the
  SNe~Ia, but SBF and SNe~Ia are not consistent with each other.
Moreover, with the revised zero point that corrects for the
metallicity dependence of the Cepheids, the other PNLF distances
published in the past decade for Fornax galaxies are $\mM =
31.10^{+0.11}_{-0.15}$ mag for NGC\,1380 \citep{feldmeier07} and $\mM
= 31.46\pm0.18$ mag for \object{NGC\,1344} \citep{teodorescu05}.  Thus,
although the PNLF method finds systematically lower distances to both
the Virgo and Fornax clusters, it is consistent with SBF in placing
NGC\,1316 well within the distance range of the other Fornax cluster
galaxies.
Further details on the SBF-PNLF comparison are given in Appendix
\ref{appenda}.

\section{SBF and integrated colours to constrain stellar population properties}
\label{sec_ssp}
 
The comparison of model predictions with observed SBF magnitudes and
SBF colours to understand the properties of the host galaxy has already
been successfully used by different authors
\citep{tal90,buzzoni93,jensen03,raimondo05,cantiello07a,buzzoni08}.

   \begin{figure*}
   \centering
   \includegraphics[width=0.96\textwidth]{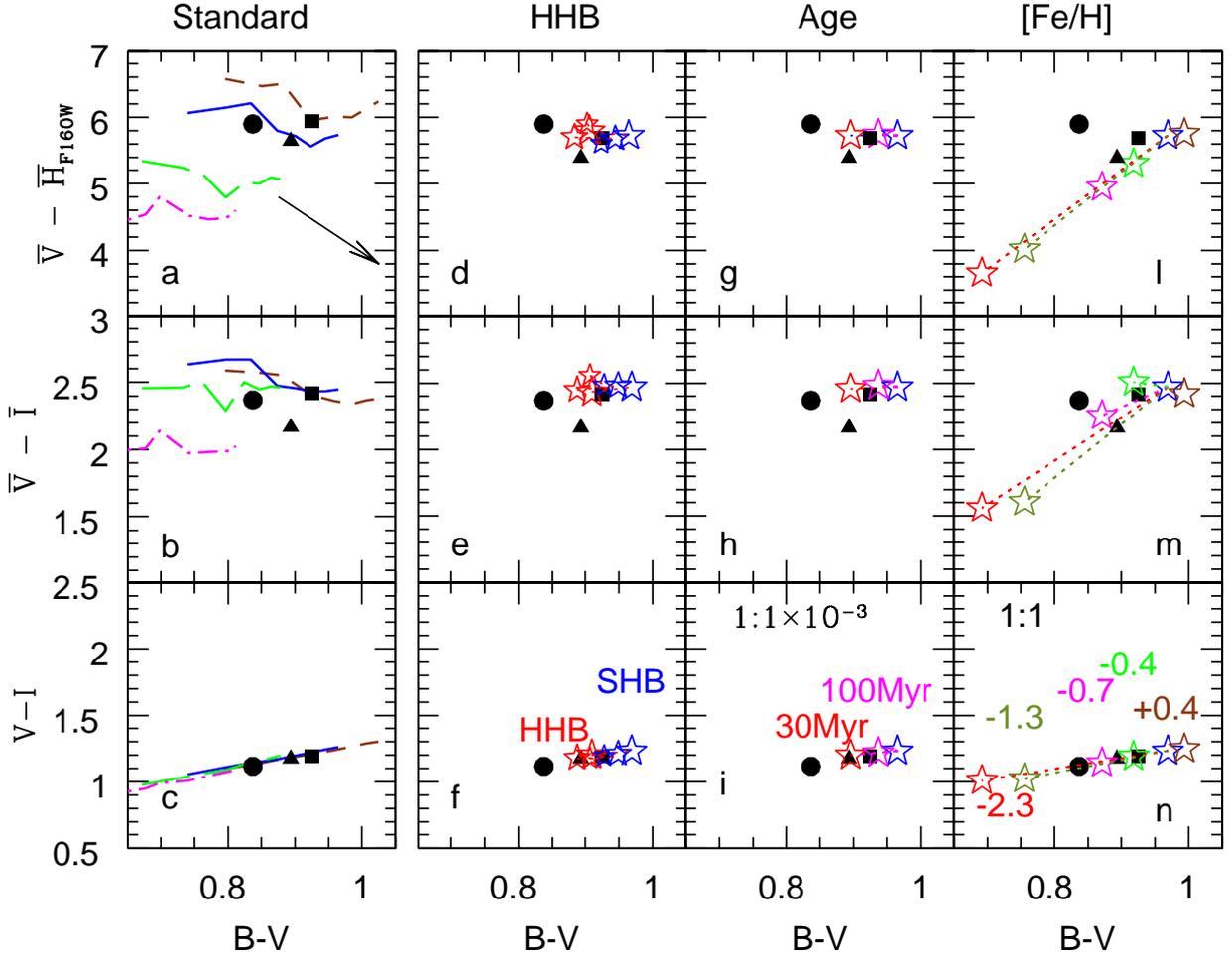}  
    \caption{Panels $a-c$: SPoT {\it standard} SSP models compared
      with SBF measurements of NGC\,1316 (full black circle). The data
      of \object{NGC\,4374} (filled square), and \object{NGC\,4621}
      (filled triangle) from \citet{cantiello11a} are also
      shown. Different line styles mark different [Fe/H] contents:
      dot-dashed (magenta), long-dashed (green), solid (blue) and
      dashed (drown) refer to [Fe/H]=$-0.7,~-0.4,~0.0$ and $+0.4$ dex,
      respectively. Models in the age-range 3--14 Gyr are shown. The
      arrow in panel $a$ indicates the direction of increasing
      age. Panels $d-n$: SPoT models obtained with non-canonical
      assumptions. In all cases the initial (reference) population has
      solar metallicity, t$\sim14$ Gyr, and is shown with a blue star,
      while the final composite population is shown with a different
      colour and connected with a dotted line to the reference
      model. The symbols for observational data are the same as in
      panels $a-c$. Panels $d-f$: A fraction equal to 50\% of the
      total HB-stars is simulated being HHB. For these models three
      ages are considered (t=10, 12 and 14 Gyr, increasing ages are
      marked with larger symbols). Panels $g-i$: A young population of
      t=30 Myr (red), or t=100 Myr (magenta) is added to the old solar
      one. The fraction in mass of old to young stars is reported in
      the lower panel. Panels $l-n$: old SSPs with various [Fe/H] are
      mixed to the solar one, as labeled. The mass fraction metal-poor
      to standard SSPs is shown in the lower panel.  \it{[See
          electronic version of the Journal for a colour version of
          the figure.]}}
         \label{ssp_a}
   \end{figure*}

   \begin{figure*}
   \centering
   \includegraphics[width=0.96\textwidth]{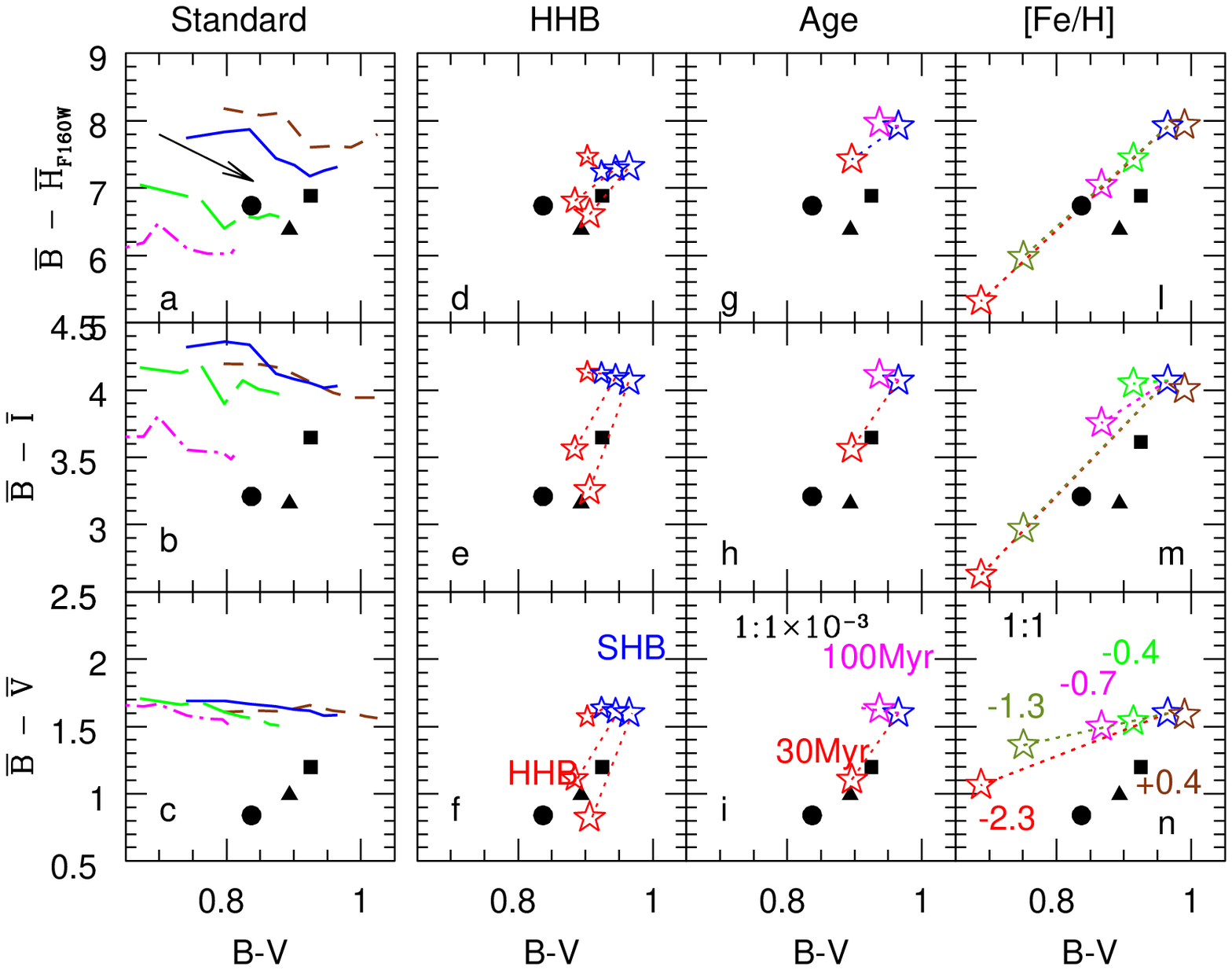}  
    \caption{Same as in Figure \ref{ssp_a}, but $\bar{B}$ is considered
      instead of $\bar{V}$. \it{[See electronic version of the Journal
          for a colour version of the figure.]}}
         \label{ssp_b}
   \end{figure*}

The SBF colour versus integrated $B{-}V$ for NGC\,1316, plus two more
Virgo cluster galaxies, NGC\,4374 and NGC\,4621 \citep[data
  from][]{cantiello11a}, are shown in Figures \ref{ssp_a} and
\ref{ssp_b}. The data in the figures are compared to SPoT SSP models
computed with standard assumptions for different metallicity and age
(panels $a-c$), while the panels from $d$ to $n$ show the predictions
for solar-metallicity standard and non-standard models: specifically,
$i)$ in panels $d-f$ SSPs with an enhanced hot horizontal branch
component (HHB, having $\sim$ 50\% stars in the canonical HB and
$\sim$ 50\% HHB stars) are considered; $ii)$ in panels $g-i$
results obtained by adding a fraction of very young stars of 30 or 100
Myr to the old, solar [Fe/H] component (in proportion 1:1000) are
plotted; and $iii)$ in panels from $l$ to $n$ predictions for models
with a mix of an old solar metallicity component, and a further one
with [Fe/H]=$-$2.3, $-$1.3, $-$0.7, $-$0.4 and +0.4 dex (in portions 1:1) are
shown.

Panel $c$ in Figure \ref{ssp_a} clearly shows the well-known
age-metallicity degeneracy that affects classical integrated colours
\citep[e.g.,][]{worthey94}. The position of galaxies in this panel
overlaps nicely with SSP models, though nothing can be said about the
stellar content. Because of the overlap between models, the field
stellar component in NGC\,1316 could be either older and more metal
poor or younger and more metal rich than the other two galaxies. In
contrast, the SBF colour versus $B{-}V$ models shown in panels $a-b$
are much less affected by the degeneracy, especially for the
$\bar{V}-\bar{H}_{F160W}$ colour. The data for the three galaxies lie
near the region of solar metallicity models. A possible interpretation
of the relative positions of the three galaxies in the
$\bar{V}-\bar{H}_{F160W}$ and $\bar{V}-\bar{I}$ versus $(B{-}V)$
colour planes, is that NGC\,1316 hosts a field component that is as
metal rich as in NGC\,4621, with [Fe/H]$\sim$0.0 dex, but
significantly younger ($3\leq t~(Gyr) \leq 10$). In fact, while both
the galaxies in Virgo lie near the edge of old SSP models, this is not
the case for NGC\,1316, for which the stellar light is polluted by an
intermediate-age component.  This is seen more clearly in some of the
other panels of the figure.

Inspecting panels $l-n$ we find that, while the colour-colour
panel $n$ shows the expected age-metallicity degeneracy, the
situation changes for SBF colours (panels $l-m$). The Virgo cluster
galaxies, in fact, overlap with the region of models obtained with the
mixing of old t=14 Gyr SSPs with different [Fe/H]s. In both the $l$ and
$m$ panels NGC\,1316 lies above the line of mixed old SSPs, suggesting
that a younger SSP is necessary to obtain a good match with the
models. Again, this is not surprising since NGC\,1316 is a known
example of an intermediate-age merger remnant, and also has
Mg$_2\sim0.25$ mag, while the other two targets have Mg$_2>0.28$~mag.

The data to models comparison is less straightforward when $B$-band
SBF magnitudes are considered. As discussed by other authors
\citep[e.g.,][]{worthey93b,cantiello07a}, $B$-band SBF cannot be used
to get reliable galaxy distances, both because the amplitude of the
signal is 2-3 orders of magnitude fainter than in optical/near-IR
bands, and because of the strong sensitivity to stellar population
properties.  This is depicted in Figure \ref{ssp_b} (panels $a-c$),
where we plot SBF colours obtained with $\bar{B}$ versus the $B{-}V$
for the three galaxies, and the standard SPoT models. One major
difference with the results in Figure \ref{ssp_a} is the substantial
mismatch between data and models seen in panel $b$ and, especially, in
panel $c$. The $\bar{B}{-}\bar{V}$ versus $B{-}V$ models are clearly
affected by the age-metallicity degeneracy, but in this case the data
are offset by $\sim 0.6$ mag with respect to the models. The mismatch
is less evident for the $\bar{B}{-}\bar{H}_{F160W}$ colour, due to the
much larger baseline of this colour.

If non-standard SPoT SSP models are taken into account (panels from
$d$ to $n$ in Figure \ref{ssp_b}) we notice that:

\begin{itemize}

\item models with an enhanced number of hot HB stars provide a good match
  to the data; 

\item the mean metallicity of the dominant stellar component in
  NGC\,1316 is apparently lower than that of the two Virgo
  galaxies. The models shown in panels $d-f$, in fact, are obtained
  assuming HHB ``enhancement'' for three different ages and solar
  metallicity. Hence, taking into account SSP models with lower
  [Fe/H], i.e., bluer colours, the matching with the position of
  NGC\,1316 will improve;

\item the presence of a diffuse very young stellar component, with an
  age of 30 Myr, seems to reduce the mismatch with data (panels
  $g-i$). However, even in the case of the merger remnant NGC\,1316,
  the presence of such a young diffuse stellar component is unlikely 
even if the presence of a relatively young stellar population is
expected in strong radio-emitter galaxies \citep{dellavalle03};

\item the improved data and models matching in panel $l$ seems to
  support the possibility of an old [Fe/H] mixed stellar
  component. However, this is ruled out by the
  comparison in panel $n$, showing a poor match to the
  $\bar{B}{-}\bar{V}$ predictions for the same data set.

\end{itemize}

At this stage it is useful to recall that the HHB scenario is
supported by other independent observations. First, we mention the
results by \citet{brown00} and \citet{brown08} who found a significant
fraction of HHB stars in M\,32 using HST $UV$ data. Second, the
puzzling presence of a strong $UV$ emission in some regular
early-type galaxies, discovered several decades ago
\citep{code72,bertola80}, is now widely interpreted as the presence of
an old hot stellar component. Although the mechanisms regulating this
component are not well understood \citep{park97,kaviraj07,han08}, some
of these old hot stellar sources may have effects on $\bar{B}$, as
independently predicted by various SSP models
\citep{worthey93b,cantiello03}, or based on empirical evidence
\citep{shopbell93,sodemann96,cantiello07a}.

We emphasize that, when computing the SBF calibration equations using
the models with HHB, we find that $\bar{B}$ brightens by $\sim$0.2
mag at fixed colour. The effect is $\lsim0.05$ mag in other bands
(negligible redwards of $z$). Such behavior, again, highlights the
uselessness of $\bar{B}$ for distance determinations, and confirms the
interest on SBF in blue bands for analyzing the properties of
unresolved blue hot stellar components.
 
In conclusion, the present analysis of the stellar population
properties for NGC\,1316 seems to confirm the results of NGC\,4374 and
NGC\,4621, i.e., that a diffuse component of hot old stars 
contributes to the SBF signal in the $B$-band. Furthermore, the
relative comparison of SBF and colour data of NGC\,1316 with those of
the two galaxies in Virgo seems to indicate that the dominant stellar
component in NGC\,1316 is younger and slightly less metal rich, as
expected for an intermediate-age merger remnant.


\section{Summary}
\label{sec_conclu}

We have measured SBF magnitudes in NGC\,1316, the brightest galaxy in
the Fornax cluster, using ground-based VLT/FORS1 data in
$BVI$-bands, and space-based $g_{F475W}$ and $z_{F850LP}$ from
ACS/WFC, plus $J_{F110W}$ and $H_{F160W}$ WFC3/IR observations.

The distance of NGC\,1316 is particularly interesting in the
  context of the cosmological distance scale. The Fornax
cluster, in fact, is the second largest cluster of galaxies within
$\lsim25$ Mpc after the Virgo cluster. However, in contrast to Virgo, the
line-of-sight depth of Fornax is small, enabling accurate calibration
of distances without the additional scatter intrinsic to the spatial
extent of the cluster.  Furthermore, NGC\,1316 is among the galaxies
with the largest number of detected Type Ia supernovae (SN\,1980N,
SN\,1981D, SN\,2006dd and SN\,2006mr). For this reason, it is a unique place
to test the consistency of SNe~Ia distances, both internally and
against other distance indicators.

Using our SBF measurements in $VIz_{F850LP}H_{F160W}$ and available
{\it empirical} calibration of the absolute $\bar{M}$, we obtained a
weighted mean distance modulus to NGC\,1316 $(\bar{m}-\bar{M})_{empirical}=31.59\pm0.05({\rm
  stat.})\pm0.20({\rm sys.})$ mag.

Additionally, we obtained SBF distances from
$VIz_{F850LP}J_{F110W}H_{F160W}$ data based on {\it theoretical}
calibrations derived from the SPoT SSP models. The resulting weighted
mean distance modulus is $(\bar{m}-\bar{M})_{theoretical}=31.60\pm0.11
({\rm stat.})\pm0.20({\rm sys.})$ mag.

The good agreement between SBF distances obtained from empirical and
theoretical calibrations is a notable result. The two classes of
calibration are, in fact, completely independent of each other, one
relying on the first two rungs of the cosmic distance scale, the
other on the present knowledge of the various ingredients that go into
stellar population synthesis (stellar evolution theory, stellar
atmospheres, etc.). Furthermore, since SBF magnitudes over such a wide
range of wavelengths depend on the properties of stars in different
evolutionary stages, even the agreement between distance moduli
obtained in different bands should be regarded as a remarkable result
particularly in terms of the high degree of reliability reached by SSP
models.

By combining the distance moduli from the two types of calibrations, we
obtain $(\bar{m}-\bar{M})=31.59\pm0.05({\rm stat.})\pm0.14({\rm sys.})$ mag, or
$d=20.8\pm0.5({\rm stat.})\pm1.5({\rm sys.})$ Mpc.

This distance modulus agrees generally well with estimates obtained
from other indicators, and with SNe~Ia light-curves analysis obtained
before 2010. A non-negligible difference exists with the most recent
analysis based on SNe~Ia by \citet{stritzinger10}, who obtained a best
estimate for the distance $\sim 17\%$ smaller than ours. The
  possible sources of the disagreement may be related to the complex
  issue of the internal extinction, and to zero point calibration
  issues of both distance indicators.

The comparison to PNLF is also subject to the lingering problem of
zero point calibration. When placed in a Cepheid distance scale
  consistent with ours, the PNLF distance to NGC\,1316 is
  $d=18.4\pm1.0 ({\rm stat.})\pm1.5 ({\rm sys.})$ Mpc and agrees
  within the quoted errors with our SBF distance. However, the
  difference between SBF and (the updated) PNLF distance remains
  non-negligible, and lowering the uncertainties (especially
  systematic; see Appendix \ref{appenda}) would be a desirable result
  for both indicators. \\

In order to analyze the properties of the dominant stellar component
in the galaxy, we compared SBF colours and integrated colours to SSP
model predictions. We found that the stellar light of the galaxy seems to be
dominated by a [Fe/H]$\lsim0.0$ dex, intermediate age stellar
population.  The comparison with analogous measurements for NGC\,4374
and NGC\,4621 also supports a scenario in which field stars in
NGC\,1316 have a younger age and slightly lower metallicity than the
two bright Virgo cluster members.

Moreover, we found that SBF predictions from standard SSP models do
not match with observations if $B$-band SBF data are included in the 
model comparison.  As in the cases of NGC\,4374 and NGC\,4612,
which showed a similar mismatch to models, we used the SPoT stellar
population synthesis code to generate SSP models with non-canonical
properties. In particular, we considered the following three cases:
starting from an old $t\sim14$ Gyr population with solar metallicity
we have $1)$ enhanced the content of hot HB stars, $2)$ added a very
young diffuse secondary component, and $3)$ added a more metal poor
SSP. As in the previous case \citep{cantiello11a}, the simulations
seem to favor the HHB component scenario.  Assuming a contribution to
$\bar{B}$ from such hot HB stars removes the discrepancy between the data
and models in this band, yet has negligible effect on SBF in other bands,
i.e., it does not affect the theoretical calibrations used to obtain
distances. \\

Our results on the distance and stellar population properties of
NGC\,1316 based on SBF analysis have shown that, despite the great
progress in recent years, many issues remain open on both
topics. Concerning distances, the calibration of distance indicators,
and the treatment of error propagation in the distance scale, still
need to be accurately and consistently analyzed. Concerning stellar
populations, SBF colours, as a new and independent stellar population
analysis technique, seem to provide useful constraints to the
properties of field stars, hidden to many classical photometric
indicators.

\newpage

 \begin{table*}
 \tiny
    \begin{center}
   \caption[]{Quality control statistics for the exposures.}
   \label{QC}
   \begin{tabular}{lllll}
         \hline\hline	

         \noalign{\smallskip}
         Name & Background  & Background & seeing & Used Data \\ 
         -	& (counts) & ($mag/arcsec^2$) & (") & - \\
         \hline
 \multicolumn{5}{c}{$B$ band} \\
 FORS.1999-12-28T02:42:27.813.fits & 888.0 & 22.25 & 0.86 & yes \\
 FORS.1999-12-28T02:54:22.346.fits & 907.4 & 22.22 & 0.81 & yes \\
 FORS.2000-01-09T02:41:46.659.fits & 983.8 & 22.13 & 1.07 & yes \\
 FORS.2000-01-09T02:52:43.240.fits & 998.9 & 22.12 & 1.01 & yes \\
 FORS.2000-01-12T01:11:59.202.fits & 1556.3 & 21.64 & 1.06 & yes \\
 FORS.2000-01-12T01:22:57.678.fits & 1527.3 & 21.66 & 1.12 & yes \\
 FORS.2000-01-13T00:56:41.525.fits & 2319.5 & 21.20 & 1.20 & yes \\
 FORS.2000-01-13T01:07:40.539.fits & 2236.8 & 21.24 & 1.14 & yes \\
 FORS.2000-01-14T03:50:58.405.fits & 1958.7 & 21.39 & 1.15 & yes \\
 FORS.2000-01-14T04:01:58.759.fits & 1580.1 & 21.62 & 1.19 & yes \\
 FORS.2000-01-15T03:10:31.420.fits & 7298.6 & 19.96 & 0.77 & no \\
 FORS.2000-01-15T03:21:27.996.fits & 7231.2 & 19.97 & 0.86 & no \\
 FORS.2000-01-17T03:28:21.613.fits & 11877.6 & 19.43 & 1.36 & no  \\
 FORS.2000-01-17T03:39:19.186.fits & 12162.4 & 19.40 & 1.30 & no  \\
 FORS.2000-01-18T01:20:21.447.fits & 12832.7 & 19.35 & 0.87 & no \\
 FORS.2000-01-18T01:31:20.684.fits & 12983.9 & 19.33 & 0.77 & no  \\
 FORS.2000-01-19T01:30:49.601.fits & 20047.5 & 18.86 & 1.38 & no  \\
 FORS.2000-01-19T01:41:43.505.fits & 21026.0 & 18.81 & 1.31 & no  \\
 FORS.2000-01-20T02:36:07.201.fits & 32829.0 & 18.33 & 1.15 & no  \\
 FORS.2000-01-20T02:47:02.044.fits & 37640.4 & 18.18 & 0.70 & no  \\
 FORS.2000-01-21T01:58:13.531.fits & 35260.4 & 18.25 & 0.70 & no  \\
 FORS.2000-01-21T02:09:11.551.fits & 35681.1 & 18.24 & 0.70 & no  \\
         \hline
 \multicolumn{5}{c}{$V$ band} \\
 FORS.2000-01-09T03:28:34.384.fits & 3261.5 & 21.13 & 0.91  & yes\\
 FORS.2000-01-09T03:39:32.352.fits & 3333.3 & 21.10 & 1.10 & yes \\
 FORS.2000-01-13T01:20:04.617.fits & 10012.2 & 19.91 & 0.98 & yes \\
 FORS.2000-01-13T01:31:04.242.fits & 69.2 & 19.93 & 0.84 & yes \\
 FORS.2000-01-13T01:39:32.083.fits & 5141.9 & 20.63 & 1.03 & yes \\
 FORS.2000-01-13T01:50:31.241.fits & 5130.5 & 20.64 & 1.09  & yes\\
 FORS.2000-01-17T04:16:19.860.fits & 24234.6 & 18.95 & 1.51 & no \\
 FORS.2000-01-17T04:27:17.584.fits & 25234.8 & 18.91 & 1.59 & no  \\
 FORS.2000-01-18T02:06:59.548.fits & 25642.0 & 18.89 & 0.76 & no  \\
 FORS.2000-01-18T02:17:56.431.fits & 23947.3 & 18.96 & 0.93 & no  \\
 FORS.2000-01-19T01:07:22.117.fits & 38819.6 & 18.44 & 1.21 & no  \\
 FORS.2000-01-19T01:18:19.487.fits & 37909.1 & 18.46 & 1.28 & no  \\
 FORS.2000-01-20T01:52:46.180.fits & 22531.1 & 18.28 & 1.25 & no  \\
 FORS.2000-01-20T01:58:44.418.fits & 29233.8 & 17.99 & 1.36 & no  \\
 FORS.2000-01-20T02:05:45.069.fits & 39325.9 & 17.67 & 0.95 & no  \\
 FORS.2000-01-20T02:11:42.821.fits & 38475.5 & 17.70 & 0.76 & no  \\
 FORS.2000-01-21T00:54:09.577.fits & 6202.4 & 17.93 & 1.14 & no \\
 FORS.2000-01-21T01:06:08.681.fits & 5118.1 & 18.14 & 1.06 & no  \\
 FORS.2000-01-21T01:30:59.576.fits & 16978.9 & 18.14 & 1.09 & no  \\
 FORS.2000-01-21T01:35:17.720.fits & 17047.7 & 18.14 & 1.11 & no  \\
 FORS.2000-01-21T01:39:35.548.fits & 15737.2 & 18.23 & 0.98 & no  \\
 FORS.2000-01-21T01:43:50.433.fits & 15366.4 & 18.25 & 1.03 & no  \\
 FORS.2000-01-21T01:48:07.248.fits & 15653.2 & 18.23 & 1.06 & no  \\
 FORS.2000-01-21T01:52:24.597.fits & 16583.5 & 18.17 & 1.08 & no  \\
             \hline
 \multicolumn{5}{c}{$I$ band} \\
 FORS.1999-12-27T03:46:00.987.fits & 9675.7 & 19.07 & 0.69 & yes \\
 FORS.1999-12-27T03:57:55.778.fits & 10469.2 & 18.99 & 0.78 & yes \\
 FORS.2000-01-09T03:05:09.924.fits & 8138.7 & 19.26 & 0.87 & yes \\
 FORS.2000-01-09T03:16:07.432.fits & 8755.6 & 19.18 & 0.82 & yes \\
 FORS.2000-01-13T02:03:14.285.fits & 9061.0 & 19.14 & 0.93 & yes \\
 FORS.2000-01-13T02:14:12.265.fits & 9847.6 & 19.05 & 0.84 & yes \\
 FORS.2000-01-15T03:34:22.320.fits & 19878.0 & 18.29 & 0.79 & yes \\
 FORS.2000-01-15T03:45:21.020.fits & 20169.8 & 18.27 & 1.00 & yes \\
 FORS.2000-01-17T03:52:07.180.fits & 18776.9 & 18.35 & 1.22 & no \\
 FORS.2000-01-17T04:03:05.404.fits & 19359.7 & 18.32 & 1.36 & no  \\
 FORS.2000-01-18T01:43:30.890.fits & 19114.6 & 18.33 & 0.65 & no  \\
 FORS.2000-01-18T01:54:29.531.fits & 19684.5 & 18.30 & 0.74 & no  \\
 FORS.2000-01-19T00:44:09.869.fits & 27488.0 & 17.94 & 1.13 & no  \\
 FORS.2000-01-19T00:55:07.408.fits & 28624.7 & 17.89 & 1.09 & no  \\
 FORS.2000-01-21T02:22:08.776.fits & 16986.9 & 17.71 & 1.01 & no  \\
 FORS.2000-01-21T02:28:02.578.fits & 17131.9 & 17.70 & 1.05  & no \\
 FORS.2000-01-21T02:33:59.741.fits & 16976.3 & 17.71 & 0.99  & no \\
 FORS.2000-01-21T02:39:57.532.fits & 16960.7 & 17.71 & 1.08  & no \\
 FORS.2000-02-04T02:06:34.213.fits & 9268.9 & 19.12 & 1.03 & yes \\
 FORS.2000-02-04T02:17:28.971.fits & 8959.7 & 19.16 & 0.85 & yes  \\

              \noalign{\smallskip}
             \hline\\
         \end{tabular}
         \end{center}

    \end{table*}

\begin{acknowledgements}
Part of this work was supported by PRIN-INAF 2010 (P.I.:
G. Clementini), PRIN-INAF 2011 (P.I. A. Grado) and PRIN-INAF 2011
(P.I.: G. Marconi), and the FIRB-MIUR 2008 (P.I. G. Imbriani).  We
are grateful to M. Capaccioli, E. Di Carlo, and I. Biscardi for useful
discussions related to this work.

This research has made use of the NASA/IPAC Extragalactic Data-base
(NED) which is operated by the Jet Propulsion Laboratory, California
Institute of Technology, under contract with the National Aeronautics
and Space Administration. This research has also made use of the
SIMBAD database, operated at CDS, Strasbourg, France, and of the
HyperLeda database (http://leda.univ-lyon1.fr).

\end{acknowledgements}

\begin{appendix}
\section{Some musings on PNLF and SBF distances}
\label{appenda}
One of the most intriguing issues in the extragalactic distance scale
is the $\sim+0.3$ mag average offset between the PNLF and SBF distance
moduli \citep{ciardullo12}.  Locally, there appears to be very little
offset, but the discrepancy increases with distance, such that the
PNLF method gives smaller mean distances for the Virgo and Fornax
clusters, as well as a smaller relative distance of Fornax with
respect to Virgo \citep[see][]{villegas10}.  Consistently,
\citet{feldmeier07} found that the value of $H_0$ obtained from SNe~Ia
was 10\% higher when the SN~Ia distances were calibrated via PNLF
distances, as compared to the $H_0$ obtained by calibrating the SN~Ia
distances by either SBF or directly from Cepheid distances.

As discussed in detail in Section \S \ref{sec_distances} and \S
\ref{sec_compare}, to reliably compare two or more distance
indicators, it is of paramount importance to verify the consistency of
the calibrations (or calibrators) used.  Both PNLF and SBF
calibrations are tied to the same primary indicator, the
period-luminosity relation of Cepheids, and to the same dataset,
i.e., the Cepheids from \citet{freedman01}. However, the zero points of
present SBF calibrations are tied to the Cepheid distances obtained
with metallicity-dependent PL relations \citep[\dplz hereafter;][this
  paper]{mei07xiii,blake09,blake10b}, while the standard PNLF
calibration relies on Cepheid distances with no dependence on
metallicity \citep[\dpl;][]{ciardullo02,feldmeier07,ciardullo12}.

In this appendix, we take the detailed discussion presented in
\citet{ciardullo02} -- who found the $\sim0.3$ mag mismatch between
the two distance indicators -- and analyze the correction terms needed
to homogenize the comparison.

First, as discussed above, \citet{tonry01} distance moduli should be
revised downward by 0.06 mag when using the \dplz.  For PNLF, instead,
\citet{ciardullo12} reported a +0.07 mag correction to the PNLF
distance moduli (0.07 mag brighter zeropoint, $M^*$) when the \dplz
distances are used instead of \dpl. It is useful to note that the
author also finds that the best-fit value to $M^*$ calibrated against
the RGB-Tip distances, i.e., independently from Cepheid distances, is
again +0.07 mag brighter than the PNLF calibration obtained from \dpl.

The two corrections: $a) -$0.06 mag for SBF distances, and $b)$ +0.07
mag for PNLF, both deriving from the adoption of the \dplz, can
justify $\sim0.13$ mag of the PNLF-SBF offset.

\begin{figure}
   \centering
   \includegraphics[width=0.48\textwidth]{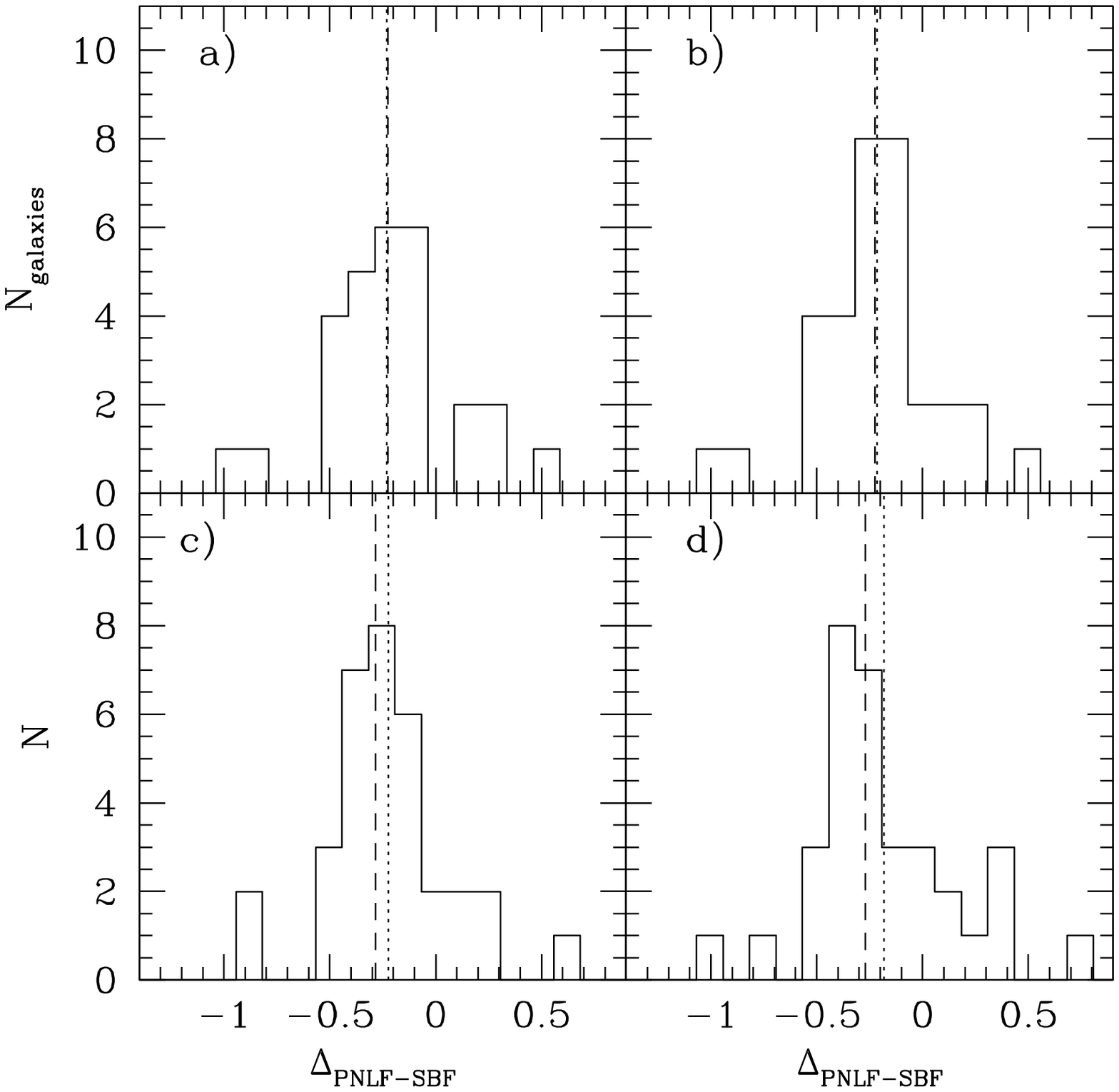} \\
   \caption{Histograms of the difference between PNLF and SBF distance
     moduli. Each panel shows different comparison/calibration
     samples. Panel $a)$: Original sample of 28 galaxies with PNLF and
     SBF distances from \citet{ciardullo02} and \citet{tonry01}
     respectively, with revised Cepheids calibration. Dotted/dashed
     vertical lines show the mean/median of the difference. Panel $b)$
     : As in $a)$ but 5 more galaxies with PNLF measurements made
     after \citet{ciardullo02} are added to the sample. Panel $c)$: As
     $b)$, but more recent ACSFCS and ACSVCS SBF distances are used
     when available. Panel $d)$: As $c)$, but the correction from
     eq. A1 by \citet{blake10b} is included to \citet{tonry01}
     distances.}
   \label{sbfpnlf}
   \end{figure}

Figure \ref{sbfpnlf} shows the histogram of the differences between
PNLF and SBF distance moduli based on various SBF and PNLF distance
and/or calibration. Panel $a)$ in the figure shows the PNLF to SBF
$\mM$ difference, $\Delta_{PNLF-SBF}$, using the original sample of 28
galaxies by \citet{ciardullo02}, with updated zero points for both
distance indicators. The mean is $\Delta_{PNLF-SBF}=-0.23$ mag, to be
compared to $\Delta_{PNLF-SBF}=-0.36$ mag before zero-point
correction.

After the \citet{ciardullo02} paper, few PNLF distances have been
obtained for galaxies with SBF measurements. Panel $b)$ in Figure
\ref{sbfpnlf} shows the PNLF to SBF difference for a total of 33
galaxies, including the distances obtained after 2002. The comparison
shown in panel $c)$ is obtained using the recent SBF distances from
the ACSVCS and ACSFCS surveys (when available) in place of the
\citet{tonry01} distances. Finally, panel $d)$ uses the same SBF and
PNLF distances of panel $c)$ except that for the galaxies with old
\citet{tonry01} corrected distances we include the further
``Q-correction'' term using eq. A1 from \citet{blake10b}. Table
\ref{tab_sbfpnlf} presents the average and median differences for all
assumptions shown in Figure \ref{sbfpnlf}.

It is worth mentioning that, according to \citet{ciardullo02}, to
properly compare PNLF and SBF, the latter distance moduli should be
further reduced by 0.04 mag. If one includes this correction term, all
differences reported in Table \ref{tab_sbfpnlf} becomes smaller, with
the best sample (i.e., $d$ in the table) providing
$\Delta_{PNLF-SBF}=-0.14\pm0.06$ mag, and a median of $-0.22$ mag.

Taking into account all the corrections described above, the offset
between SBF and PNLF distances is reduced to half of the original
estimate, once the proper calibrations and recent/updated
distances are used for both indicators. However, even in the best case
there is a non-negligible $\lsim0.2$ mag offset between SBF and PNLF
that implies $\sim10\%$ larger SBF distances.
Again, this difference occurs mainly beyond $\sim10$ Mpc, and is
similar to the 10\% larger $H_0$ obtained when calibrating SNe~Ia via
PNLF instead of Cepheids \citep{feldmeier07}.

Even though the data presented in Table \ref{tab_sbfpnlf} (with the
possible further +0.04 mag improvement cited above) suggest that the
best average difference is statistically consistent with zero --
especially taking into account the systematic uncertainties, not
considered in this comparison -- the scatter we find is larger, or
nearly equal to the squared sum of the estimated internal scatters of
both indicators. Taken at face value, this result either means that
the internal scatter of one or both indicators is underestimated, or
that a real systematic offset exists between the two.

As another test, to further check the latter concern, we analyzed the
PNLF to SBF offset by considering late-type and early-type galaxies
separately. Using sample $d$ in Table \ref{tab_sbfpnlf}, we find the
following differences: $\Delta_{PNLF-SBF}^{Early}=-0.23\pm0.05$ mag
(median $-0.30$ mag) based on the data of 24 galaxies, and
$\Delta_{PNLF-SBF}^{Late}=0.04\pm0.17$ mag (median $-0.02$ mag) for
the remaining 9 galaxies. The result shows that for the class of
galaxies used to derive SBF and PNLF zero points, i.e., the late types
hosting Cepheids, there is no statistically significant offset between
the two indicators. On the other hand, for early-type galaxies the
offset is large, and statistically inconsistent with zero.  Related
to this, one fundamental difference between the two indicators is that
absolute SBF magnitudes are ``corrected'' for galaxy stellar content,
i.e.,  the difference between early- and late-type galaxies is taken into
account with SBF, while the PNLF distances are based on the constancy
of $M^*$ for both types of galaxy.

The calibration of SBF magnitudes, and its dependence on galaxy type,
has been analyzed in detail over a ground-based sample of $\sim$300
galaxies by \citet{tonry01}, and more recently from HST data of
$\sim150$ galaxies by \citet{blake09,blake10b}. The derived SBF
calibrations, as is well-known, include a colour-dependent term,
which, as also shown by SSP models
\citep{worthey93a,cantiello03,raimondo05}, is basically a metallicity
correction term. In optical bands, this correction term implies
fainter SBF magnitudes for redder/more metal-rich systems.

In contrast, the $M^*$ calibration to PNLF does not include any
metallicity dependent term for bright galaxies. A dependence of the
PNLF $M^*$ to metallicity has been found by \citet{ciardullo02} and
\citet{ciardullo12}, however the authors conclude that such dependence
is relevant only in small, metal-poor systems.

\begin{figure}
   \centering
   \includegraphics[width=0.48\textwidth]{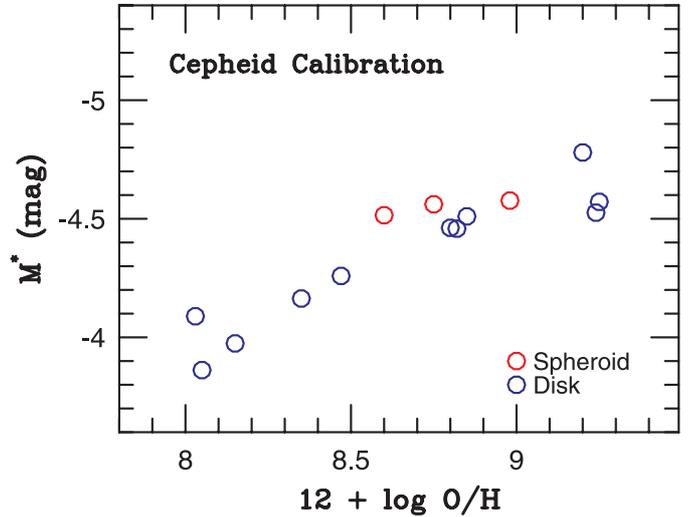} \\
   \caption{Figure 5 from \citet{ciardullo12} showing the sole data
     with metallicity corrected Cepheids distances. \it{[See electronic version of the
          Journal for a colour version of the figure.]}}
   \label{pnlfz}
   \end{figure}

Inspecting the open circles in Figure 5 of \citet[open circles mark
  the data obtained from \dplz distances]{ciardullo12}, reported in
Figure \ref{pnlfz}, one can see that $a)$ the trend of $M^*$ with
metallicity has the opposite sign with respect to SBF, meaning that
$M^*$ gets brighter for more metal-rich systems, and that $b)$ some
residual correlation of $M^*$ with metallicity also appears in the
high metallicity regime.

This strongly suggests that there may be some unaccounted for residual
dependence of the PNLF calibration on the metallicity, presently
unquantified because of the relatively limited sample -- though
detectable even in present datasets (Fig. \ref{pnlfz}). If so, one
possible explanation for the irreducible $\sim-0.2$ mag offset between
PNLF and SBF is that it may be due to a bias introduced by data from
early-type galaxies, which are intrinsically on average more massive
and more metal-rich than late types and, therefore, would typically
have larger distances with respect to the current metallicity-independent
calibration of the PNLF method.

\begin{table*}
\caption{SBF to PNLF comparison.}
\label{tab_sbfpnlf}
\centering
\begin{tabular}{l c c c}
\hline\hline 
Sample                       & Number of galaxies &  $\Delta_{PNLF-SBF}$ ($rms$) &  Median difference  \\
                             &                    &      (mag)          &            (mag)    \\
\hline
Original$^\mathrm{[1]}$        &        28          &   $-$0.36  (0.31)  &    $-$0.36            \\   
a$^\mathrm{[2]}$               &        28          &   $-$0.23  (0.31)  &    $-$0.23            \\
b$^\mathrm{[3]}$               &        33          &   $-$0.21  (0.29)  &    $-$0.23            \\
c$^\mathrm{[4]}$               &        33          &   $-$0.22  (0.30)  &    $-$0.29            \\
d$^\mathrm{[5]}$               &        33          &   $-$0.18  (0.34)  &    $-$0.26            \\
\hline
\end{tabular}
\begin{list}{}{}
\item [[1]] Original PNLF sample by \citet{ciardullo02} with SBF distances from \citet{tonry01}.
\item [[2]] \citet{tonry01} SBF \& \citet{ciardullo02} PNLF with revised zero points based on metallicity dependent PL relations for Cepheids (see text).
\item [[3]] \citet{tonry01} SBF \& \citet{ciardullo02} PNLF plus more
  recent PNLF distances from \citet[][NGC\,1380, \object{NGC\,4526} and revised
    NGC\,1316]{feldmeier07}, \citet[][\object{NGC\,4697}]{sambhus06},
  \citet[][NGC\,1344 and \object{NGC\,821},
    respectively]{teodorescu05,teodorescu10} and
  \citet[][\object{NGC\,4376}]{herrmann08}. For both distance indicators
  revised zero points are used, as in sample $b$ except for \object{NGC\,4697}, NGC\,1344 and
  NGC\,821 whose PNLF distance is independent from
  \citeauthor{ciardullo02} calibration.
\item [[4]] Updated SBF distances from ACSVCS \citep{cote04} \& ACSFCS
  \citep{jordan07} (when available). Old SBF and PNLF distances as in sample $b$.
\item [[5]] As sample $c$ except that SBF with Q--corrected SBF
  distances (see text) are used for the old \citet{tonry01} distance
  moduli.
\end{list}
\end{table*}

\section{Some notes on Type~Ia SNe distances from Stritzinger et al. (2010)}

\label{appendb}

To understand the possible causes of the difference between our and
Str10 distance we must recall that, as for SBF, in order to calibrate
SNe~Ia light curves one must rely on sources at known distance and/or
with well-known intrinsic properties, and then standardize the
absolute magnitude of the SN~Ia \citep[e.g.,][]{phillips93}.

Str10 derives the distance to NGC\,1316 using three different methods:
the EBV, the Tripp method, and the near-IR light-curves. The authors
also use the MLCS2k2 method on SN\,2006mr, obtaining a distance that is
50\% further than the average they derived from the normal events.

In the following we discuss each one of the three methods used by
Str10, trying to highlight the possible causes leading to the observed
difference.

\subsection{The Colour Excess Method}
Str10 adopted the calibration from \citet[][AJ in press at the time
  Str10 was published]{burns11} to obtain the distance moduli with the
``EBV'' model in their fitting package SNooPy. For this model they
adopt the recommended calibration from \citet{burns11}, which uses a
sub-sample of unreddened SNe~Ia from \citet{folatelli10}, excluding
fast declining objects, as SN\,2006mr. \citet{folatelli10}, in turn,
calibrate their dataset using 26 SNe~Ia at $z>0.01$, whose distances
are based on Hubble's law assuming $H_0=72~~km~s^{-1}~Mpc^{-1}$,
and three SNe~Ia at $z<0.01$ with direct distance measurements. The
three nearest objects include SN\,2006mr, whose host's distance is
assumed $(m-M)=31.59\pm0.08$ mag from the SBF measurements by
\citet{cantiello07b}.

The SNooPy/EBV method relies (also) on an estimate of the internal
reddening around the SN~Ia. Str10 find that all four SNe~Ia in NGC\,1316
have negligible internal extinction. The authors warn about some
complications in the interpretation of the data.  With respect to
spectroscopic analysis of Na~I D absorption in the spectra of SN\,2006dd
and SN\,2006mr, these authors state that {\it ``the very strong Na~I D
  absorption observed in SNe~Ia 2006dd and SN\,2006mr is totally
  inconsistent with the low host galaxy reddening we derive from the
  light curve observations''} (Str10, Section 4), as the colour
evolution of the two SNe~Ia closely resembles that of unreddened
SNe~Ia.  However, while there is a general observational
  agreement on higher colour excess corresponding to higher Na~I D
  equivalent width (EW), this correlation is tight in high-resolution
  spectra, but the scatter increases substantially at lower resolution
  \citep{blondin09,poznanski11,poznanski12}, with considerable
  confusion due to the blending of the Na~I D doublet. For instance,
  using Figure 5 in \citet{blondin09}, at EW$\sim1.5$ (similar to
  that obtained by Str10 for NGC\,1316) $E(B-V)_{host}$ ranges
  from $\sim0.1$ to $\sim1.5$ mag.  As a further complication, besides
  the possible line blending, the line profiles for both SNe show
  clear evidence for structure, interpreted by Str10 as evidence of
  the presence of two unresolved sodium components. Nevertheless,
  coupling the EW measurements from Str10 (their Table 10, obtained
  from the authors' highest dispersion spectra), with equations
  (7)-(8) from \citet{poznanski12}, we obtain $E(B-V)_{host}>0.45$ and
  $>0.15$ mag for SN\,2006dd and SN\,2006mr, respectively (with mean
  values $\sim 1.4$ and $\sim0.6$ mag).

Clearly, as also highlighted by Str10, the strong Na~I D absorption
associated with no internal reddening might also be indicative of a
non-standard gas-to-dust ratio for Fornax A, which given the merger
history of this galaxy does not seem unreasonable.

   \begin{figure*}
   \centering
   \includegraphics[width=0.48\textwidth]{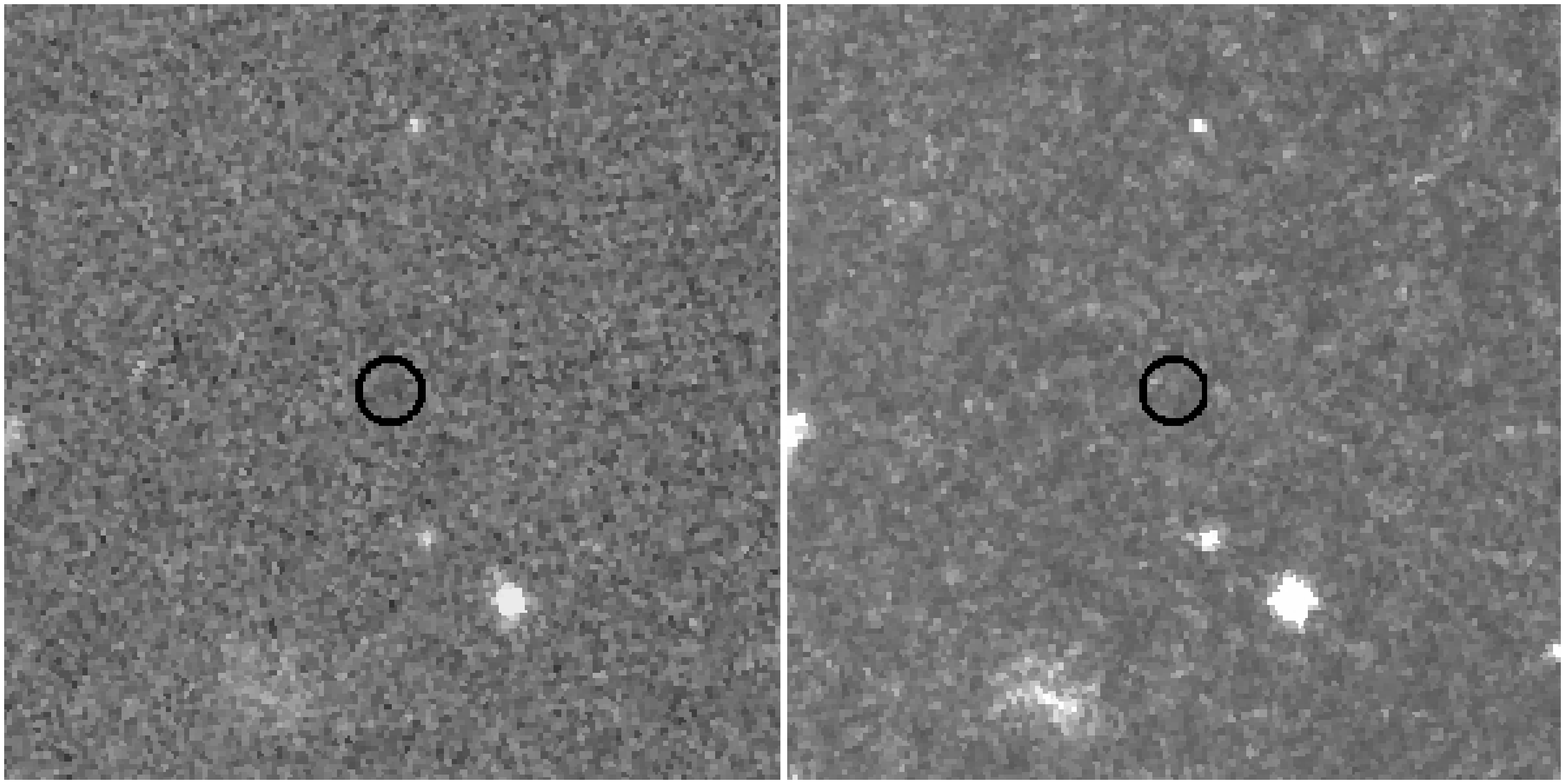} SN\,1981 D  \\
   \includegraphics[width=0.48\textwidth]{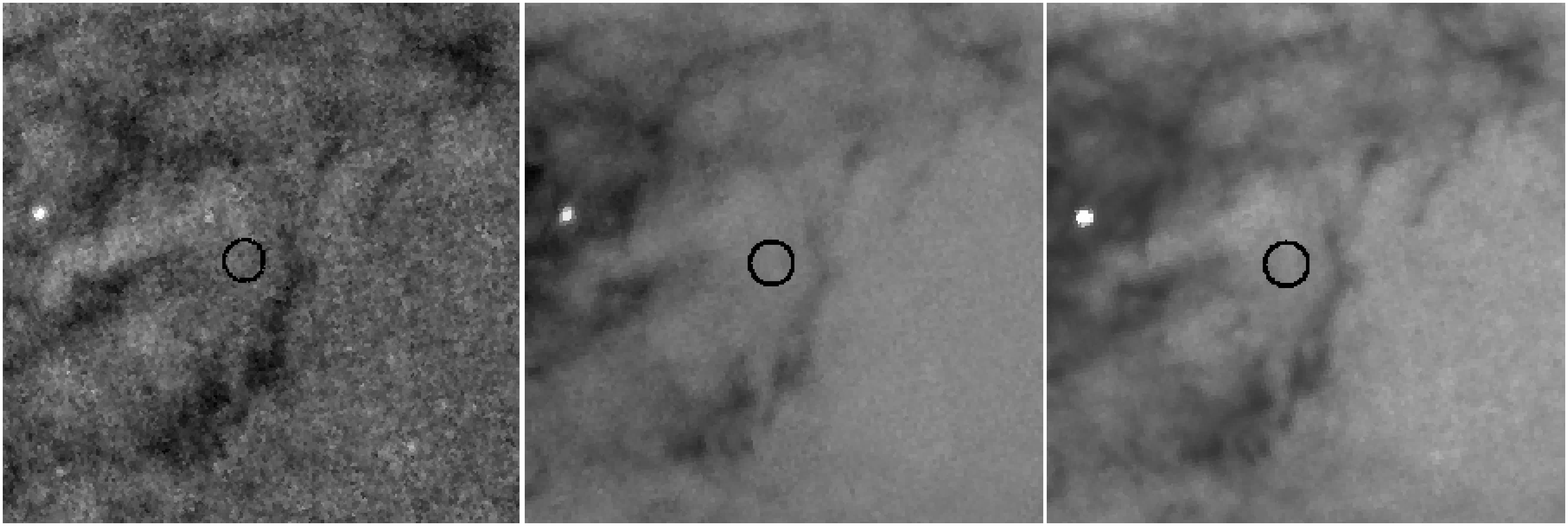} SN\,2006dd\\
   \includegraphics[width=0.48\textwidth]{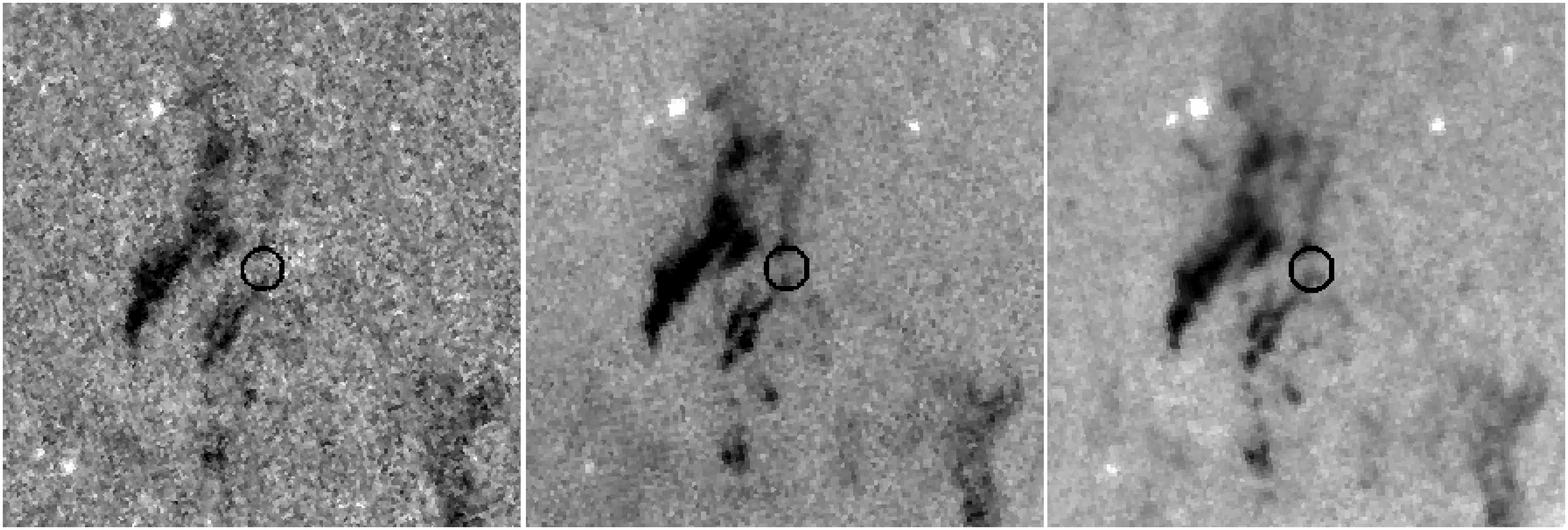} SN\,2006mr \\
  \vskip 1cm
   \includegraphics[width=0.48\textwidth]{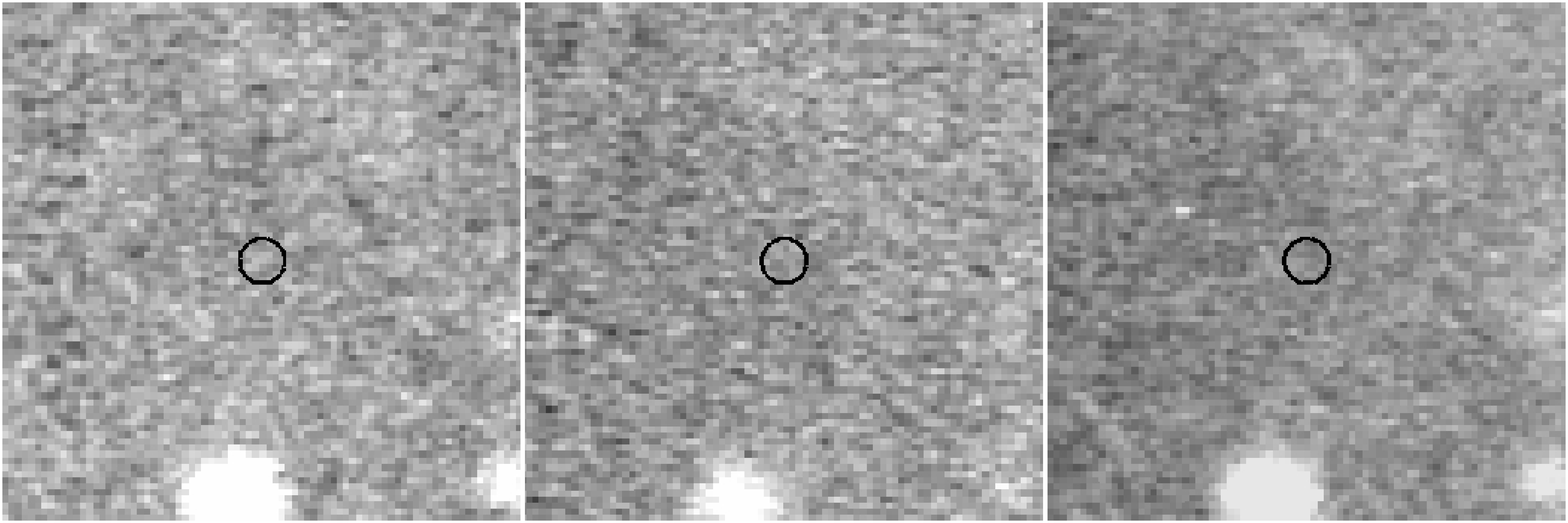} SN\,1980 N \\
   \caption{Zoom in $6\times6 arcsec^2$ of the residual frames
     centered on the positions of the four SNe~Ia, as labeled. Black
     circles mark the position of the respective SNa. UVIS/$F336W$,
     ACS/$F435W$ and ACS/$F555W$ frames (left, middle and right
     panels, respectively) are shown for the SN\,1981D, SN\,2006dd and
     SN\,2006dd. The panels showing SN\,1980N are $g$, $r,$ and $i$
     observations taken from the Gemini Telescope archive.}
   \label{sne_ubv}
   \end{figure*}

A $6\arcsec \times 6\arcsec$ zoom of the regions around the four
SNe~Ia is shown in Figure \ref{sne_ubv}. In the upper part of the
figure we show the HST UVIS/$F336W$, ACS/$F435W$ and ACS/$F555W$
residual frames of the three SNe~Ia located within the frames analyzed
in this work (SN\,1981D, SN\,2006dd, SN\,2006mr). For sake of completeness,
we have obtained archival $g$, $r$, and $i$-band Gemini/GMOS-S data of
the region around the SN\,1980N, shown in the lower panels of Figure
\ref{sne_ubv}. As evidenced in the figure, there are undeniable
patterns of dust near the position of the 2006 SNe~Ia, though one
  cannot decide whether the SNe are behind, in front of, or within such
  dust lanes.

The comparison with previous estimates of internal extinction from the
literature show that for SN\,1980N and SN\,1981D, \citet{jha07} find
$E(B{-}V)_{host}$ values larger than Str10 (three times larger in the
case of SN\,1981D) but in agreement within uncertainties with the SNooPy
fits. The agreement gets worse if the extinctions derived by Str10 from
near-IR data are taken into account. Note, however, that Str10
  corrected the optical and near-IR photometry for host galaxy
  contamination.  Such a correction, although negligible for the case of
  SN\,1980N, does not seem to be discussed by \citeauthor{jha07}

For both the two most recent SNe~Ia, 2006dd and 2006mr, \citet{maoz08}
estimate an internal extinction of $\sim$0.08 mag. In the case of SN\,2006dd,
SNooPy provides $E(B{-}V)_{host}=0.043\pm0.008$ mag; while, as mentioned
above, SNooPy cannot be used for fitting the light-curves of the fast
declining SN\,2006mr.

In conclusion, the three SNe~Ia used by Str10 to get the best estimate
of $(m{-}M)$ could be controversial in terms of internal extinction,
affecting both the estimate of $\mM$ with the EBV method, and the
associated uncertainties.

\subsection{The Tripp Method}
The second method adopted by Str10 is based on the two-parameter
model of \citet{tripp98} which, differently from the EBV method, can
also be applied to fast declining SNe~Ia. The calibrating sample is again
taken from \citet{folatelli10}, and SN\,2006mr is omitted in the
re-computed calibration relations, to avoid circularity.

The distance to the three normal SNe~Ia with this method is consistent
with the estimates based on the SNooPy/EBV method. In contrast, the
$(m{-}M)$ obtained with the data of SN\,2006mr is $\sim+0.5$ mag larger
than the average of the other three SNe~Ia.

One interesting point to note is that if one uses a distance modulus
of $\sim$31.2 mag to NGC\,1316, as derived by Str10 from the three
normal SNe~Ia, then the data-point of SN\,2006mr placed in Figure 16 of
\citet{folatelli10} is more than +0.5 mag off the linear relation
drawn by the authors, with a scatter to the relation much larger than
the $rms$ reported in the cited figure. Certainly, changing the
distance modulus of one of the calibrating data-points in the
\citeauthor{folatelli10} sample implies changing the linear
calibration relation shown in the cited figure, and possibly reduces
the offset between data and fit. In any case, though, using
$(m-M)\sim31.2$ entails considerably increasing the scatter in the
calibrating sample of the Canergie Supernova Project, with NGC\,1316
being one of the nearest objects and also the host galaxy with the
largest scatter.

\subsection{The near-IR method}
The last method used by Str10 is based on near-IR light-curves of
SNe~Ia, calibrated using \citet{krisciunas09} absolute near-IR peak
magnitudes without NGC\,1316 data.  The \citeauthor{krisciunas09}
calibration of near-IR peak magnitudes adopts new observations of
SNe~Ia, and data previously published by the same team
\citep{krisciunas04a,krisciunas04b}. For the nearby galaxies, the
authors adopted distances based on either SBF or Cepheids. From a
careful reading of the cited papers, we find that
\citet{krisciunas04a} obtained the $JHK$ calibration from 16
SNe~Ia. For three nearby galaxies, NGC\,1316, \object{NGC\,4526}, and
\object{NGC\,5128}, the authors adopt the SBF distance from \citet[][based
  on \dplz distances]{ajhar01}, while for \object{NGC\,4536} and \object{NGC\,3368} the
\dpl distances from Cepheids is used. Both SBF and Cepehids distances
are based on the same \citet{freedman01} calibrating sample.

\citet{krisciunas04b} extended the sample of SNe~Ia with well-sampled
near-IR light-curves to about 20 objects. The authors added two more
supernova-host galaxies with SBF distances -- \object{NGC\,4374} and \object{NGC\,3190}
-- to the previous list of nearby galaxies. However, in contrast with
\citet{krisciunas04a}, they adopted the \citet{tonry01} distance
moduli, which are based on \dpl; that is, they are 0.06 mag larger than
the $\mM$ reported by \citet{ajhar01}.

Finally, the most recent calibration of near-IR light-curves of SNe~Ia
by \citet{krisciunas09}, used the distance moduli for nearby galaxies
from the SBF survey by \citet{tonry01} with the revised
\citet{jensen03} zero points \citep[+0.16 mag with respect
  to][]{tonry01}. More specifically, in the new list of $\sim$25
supernova-host galaxies, the authors added three new objects,
\object{NGC\,936}, \object{NGC\,1201} and \object{NGC\,1371}, with SBF-based distances. In
addition, \citeauthor{krisciunas09} adopted a revised SBF distance to
NGC\,5128 from \citet{jensen03}. However, the authors corrected the
$(m{-}M)$ by +0.16 mag, which is the difference between the
\citet{tonry01} and \citet{jensen03} calibrations, and not by +0.10
mag, i.e., the difference between the \citet{ajhar01} and
\citet{jensen03} calibrations.

\begin{table*}
\caption{Nearby galaxy sample used for near--IR calibration: original and revised distances.}
\label{tab_deltas}
\centering
\begin{tabular}{l c l c c c}
\hline\hline 
Galaxy               & $\mM_{near-IR}$ &  Reference        & Method     & $\mM_{Revised}$&      $\Delta(Revised-Orig.)$ \\
\hline 
\object{NGC\,4526}            &  31.08    & \citet{krisciunas04a} & SBF        &    31.08    &       +0.00  \\
\object{NGC\,4536}            &  30.80    & \citet{krisciunas04a} & Cepheids   &    30.87    &       +0.07  \\
\object{NGC\,3368}            &  29.97    & \citet{krisciunas04a} & Cepheids   &    30.11    &       +0.14  \\
\object{NGC\,4374}            &  31.32    & \citet{krisciunas04b} & SBF        &    31.26    &       $-$0.06  \\
\object{NGC\,3190}(\object{NGC\,3226}) &  31.86    & \citet{krisciunas04b} & SBF        &    31.80    &       $-$0.06  \\
\object{NGC\,936}             &  31.65    & \citet{krisciunas09}  & SBF        &    31.75    &       +0.10  \\
\object{NGC\,1201}            &  31.37    & \citet{krisciunas09}  & SBF        &    31.47    &       +0.10  \\
\object{NGC\,1371}(Eridanus group) & 31.84& \citet{krisciunas09}  & SBF        &    31.94    &       +0.10  \\
\object{NGC\,5128}            &  27.90    & \citet{krisciunas09}  & SBF        &    28.06    &       +0.16  \\
\hline 
\multicolumn{5}{l}{Average (median) correction on the sample of nearby galaxies}    &  0.06$\pm$0.03 (0.085)            \\
\multicolumn{5}{l}{Average correction on the complete sample of galaxies (assume zero for distant galaxies)}  &  0.02$\pm$0.01          \\
\hline\hline 
\end{tabular} \\
\end{table*}

Hence, the calibration of the absolute magnitudes in near-IR bands
used by Str10 relies on a sample of $\sim$25 SNe~Ia, with nine nearby
calibrators (after excluding NGC\,1316) having distances based on
non-homogeneous calibrations. In Table \ref{tab_deltas} we summarize
the distance moduli used by \citeauthor{krisciunas09} to calibrate the
maxima of SNe~Ia light-curves, and the ones revised by us in order to
be $a)$ internally homogeneous, and $b)$ consistent with the Cepheid
distances used in this work. As shown in the table, the revised
distance moduli are on average 0.06 mag larger than the ones used for
the original SNe~Ia near-IR calibration. If one simply takes the
average of these numbers, adding a sample of 15 objects more (25 total
SNe~Ia minus the nine nearby objects and NGC\,1316) where no shift has
to be applied, the correction to the absolute magnitudes in Table 9 of
Str10 is $\sim-0.02$~mag \citep[a correction that should be applied to
  the calibration by][]{krisciunas09}.

Although a +0.02 mag shift in Str10 distance moduli goes in the
direction of reducing the mismatch between our and Str10 distances,
the amplitude of the correction is negligible. However, it
suggests that the uncertainties associated with the near-IR
calibrations might be underestimated. 
\vskip 0.3cm
\noindent
In conclusion, the analysis presented in this Appendix highlights two
main issues: $i)$ the homogeneity of the calibrators used, and $ii)$
the estimate of internal extinctions for SNe~Ia (although one should
not forget that the quality of the SN\,1980N and SN\,1981D data is lower
than others). While the first issue listed works in the direction of
reducing the difference between our and Str10 distance moduli, both
issues imply an increase of the present levels of
statistical/systematic uncertainties on SNe~Ia distances.

\end{appendix}

\newpage

\bibliography{ms_printer}
\bibliographystyle{aa}

\clearpage


\end{document}